\documentstyle[sprocl,epsfig]{article}

\font\rm=cmr8

\def\be{\begin{equation}}
\def\ee{\end{equation}}
\def\bea{\begin{eqnarray}}
\def\eea{\end{eqnarray}}

\begin{document}

\title{BOTTOM QUARK PHYSICS\\
PAST, PRESENT, FUTURE
\footnote{Talk given at ``Symposium on Probing Luminous and Dark
Matter, honoring Adrian Melissinos'', Rochester, October, 1999.}
}

\author{EDWARD H. THORNDIKE}

\address{Department of Physics, University of Rochester\\
Rochester, NY 14627, USA\\
E-mail: thorn@pas.rochester.edu} 




\maketitle\abstracts{ }

\section{Introduction}

Let me start by reminding you what's going on at all the major High Energy
Physics laboratories around the world. At CERN, the LEP program is winding
down, and the LHC (Large Hadron Collider) is the Lab's future. 7 TeV
protons on 7 TeV protons, a center of mass energy of 14 TeV. Four large
detectors are planned. Two, ATLAS and CMS, will study high $p_{_T}$
physics, searching for Higgs, SUSY, etc. One, ALICE, will collide high $Z$
nuclei (when protons aren't being collided), and study the quark-gluon
plasma. And one, LHC-B, will study bottom quark physics. It is a sobering
thought that a 14 TeV accelerator will be used to study a 5 GeV object, 3
orders of magnitude down the energy scale. (But one should not forget that
the Tevatron is used to study kaon physics, again 3 orders of magnitude
down the energy scale.)

At DESY, the main facility is HERA, an electron-proton collider, with 800
GeV protons on 30 GeV electrons. The two principal detectors, H1 and ZEUS,
study these collisions, investigating deep inelastic scattering over a
kinematic range far broader than heretofore. But the proton beam will also
be used, on a fixed target (wires in the fringe of the beam) for bottom
quark physics, in the HERA-B experiment.

At KEK, in Japan, TRISTAN, an $e^+ e^-$ collider operating at a
center-of-mass energy of 60 GeV has been shut down, and replaced by an
asymmetric $e^+ e^-$ collider, 8 GeV on 3.5 GeV, a center of mass energy of
10 GeV, to do bottom quark physics, with the Belle experiment.

At SLAC, the SLC (SLAC Linear Collider), $e^+ e^-$ collisions at
center-of-mass energies around 90 GeV, has been shut down, and replaced by
PEP-~II, an asymmetric $e^+ e^-$ collider, 9 GeV on 3 GeV, a center of mass
energy of 10 GeV, to do bottom quark physics with the BaBar experiment.
Thus, the study of the $Z^0$, a 90 GeV object, is giving way to the study
of the $b$ quark, a 5 GeV object.

At Fermilab, the main facility is the Tevatron, which collides 1 TeV
protons against 1 TeV antiprotons, for a center-of-mass energy of 2 TeV.
There are two general purpose detectors, operated by two large
collaborations, CDF and D$\emptyset$. The primary goal of the running
recently completed was the discovery of the top quark. Goals for the next
running period (Run II) include precise measurements of top quark and $W$
boson masses, and searches for ``new physics'' -- Higgs, SUSY, etc. But
CDF has had an active program in bottom quark physics, and foresees an
expanded program in Run II. A displaced vertex trigger is being
implemented, in part to strengthen the $b$ physics
program. D$\emptyset$ has done little $b$ physics
 so far, lacking a magnetic field in the central tracking volume.  They are
remedying this for Run II, and anticipate an active $b$ physics program.
And serious consideration is being given to a third detector, B-TeV, which
would be a dedicated bottom quark experiment.

Finally, Cornell's Laboratory for Nuclear Studies, with a symmetric $e^+
e^-$ collider (CESR), has been doing bottom quark physics for two decades.

So, bottom quark physics {\it must} be interesting, because {\it all} the
major labs have it as part of their program. {\it Why} is bottom quark
physics so interesting? (The cynic might argue that the labs are into
bottom quark physics because it's affordable. There is perhaps some truth in
this. But it doesn't explain 
LHC-B. It doesn't explain
the interest in bottom quark physics within CDF, nor SLAC's preference for
studying a 5 GeV object over a 90 GeV object.) Why is bottom quark physics
interesting? A primary goal of my talk will be to answer that question for
you.

Bottom quark physics can be conveniently divided into three eras;

\begin{itemize}

\item The Early Days -- 1977-88, further divided into Discovery -- 1977-80,
and Roughing out the Qualitative Features -- 1980-88

\vspace{-.1in}
\medskip
\item Beginnings of Precision Measurements and Rare Decay Studies -- 1989-98

\vspace{-.1in}
\medskip
\item The `Factory' Era -- 1999-??

\end{itemize}

In Section 2, I'll discuss the early days.

In Section 3, I'll point out a change in objective that took place around
1990, and give a brief review of the flavor sector of the Standard Model.

Then, in Sections 4, 5, and 6, I'll discuss three of the ``hot topics''
in $b$ physics today: determination of $|V_{ub}/V_{cb}|$, rare hadronic $B$
decays, and the radiative penguin decay $b \rightarrow s \gamma$.

\section{The Early Days}

\subsection{Discovery -- 1977-80}

The $b$ quark was discovered in its hidden form (``hidden beauty'',
``covered bottom'') at Fermilab, in 1977, by Leon Lederman and
collaborators. They measured the mass distribution of dimuon pairs from
collisions of 400 GeV protons  on a nuclear fixed target, and observed a
structure consisting of two or more peaks in the 9.4-10.0 GeV region (see
Fig. 1). The immediate (and correct) interpretation was a bound system of a
quark-antiquark pair, charge $-$1/3 quarks. The bound system was named the
Upsilon $(\Upsilon)$. 

\begin{figure}[t]
\begin{center} \epsfig{file=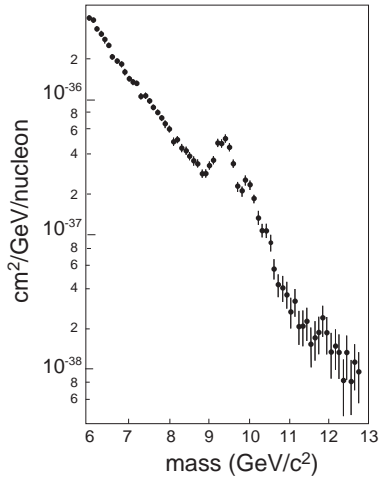, width=2in} \end{center}
\caption{Dimuon mass distribution from collisions of 400 GeV protons with a
nuclear target, showing the $\Upsilon$ states (Lederman and collaborators).}
\label{fig:Lederman}
\end{figure}

The DORIS $e^+ e^-$ storage ring at DESY, at the time of the $\Upsilon$
discovery, had insufficient energy to produce $\Upsilon$'s.  The machine
energy was increased, and in 1978, straining their RF, physicists at DORIS
observed two narrow resonances, $\Upsilon$(1S) and $\Upsilon$(2S). They
could go no higher.

The CESR $e^+ e^-$ storage ring at LNS, Cornell, gave first luminosity to
the CLEO and CUSB detectors in October, 1979. The $\Upsilon$(1S) and
$\Upsilon$(2S) resonances were quickly located, and in December, in time to
be ``added in proof'' to the Lab's Christmas card, the $\Upsilon$(3S)
was discovered (see Fig. 2).

\begin{figure}[h]
\begin{center} \epsfig{file=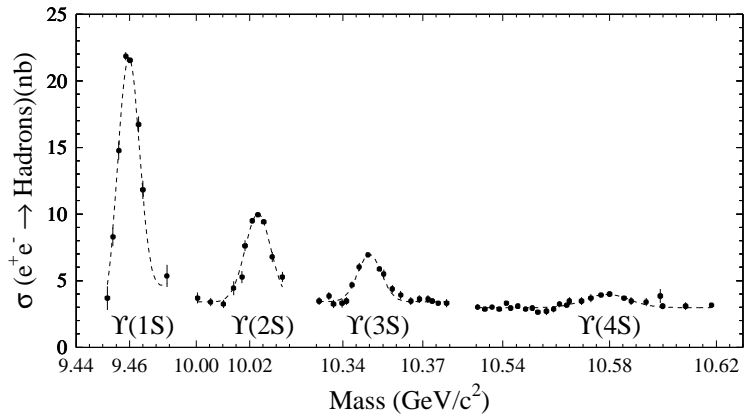, width=4in} \end{center}
\caption{Hadronic cross section vs. center-of-mass energy, showing three
narrow resonances $\Upsilon$(1S,2S,3S), and one broad resonance
$\Upsilon$(4S) (CLEO).}
\label{fig:UPS1234S}
\end{figure}

The three resonances, $\Upsilon$(1S), $\Upsilon$(2S), $\Upsilon$(3S) were
all narrow, with widths less than the instrumental resolution (beam energy
spread). The production rates, leptonic decay branching fractions, level
spacings, all matched very well with the bound
 $b \overline b$, charge $-$1/3 quark interpretation.

While there was no doubt, by then, about the {\it existence} of the bottom
quark, the studies needed to determine further properties were of its weak
decays. These could not be obtained from `hidden beauty,' because a
bound $b \overline b$ system decays via the strong interaction, with $b$
and $\overline b$ quarks annihilating each other, forming gluons or a
virtual photon. For studies of the weak decay of the bottom quark, 
``bare bottom,'' or ``naked beauty'' was needed.

Bare bottom was discovered at CESR by the spring of 1980. A scan, measuring
cross section for production of hadronic events vs center-of-mass energy,
above the $\Upsilon$(1S),
$\Upsilon$(2S), $\Upsilon$(3S), revealed another resonance. This one was
measurably broad (see Fig. 2), indicative of a rapid decay into
$b$-flavored mesons, $\Upsilon$(4S) $\rightarrow B 
\overline B$. The compelling evidence for bare bottom came from the yield
of muons and electrons, which also peaked at the $\Upsilon$(4S) resonance
(see Fig. 3), indicating the decay sequence
$\Upsilon$(4S) $\rightarrow B \overline B$ (via the strong interaction),
followed by $B 
\rightarrow X \ell V$ (via the weak interaction). Leptons, a tell-tale
signature of a weak decay, established bare bottom.

\begin{figure}[h]
\begin{center} \epsfig{file=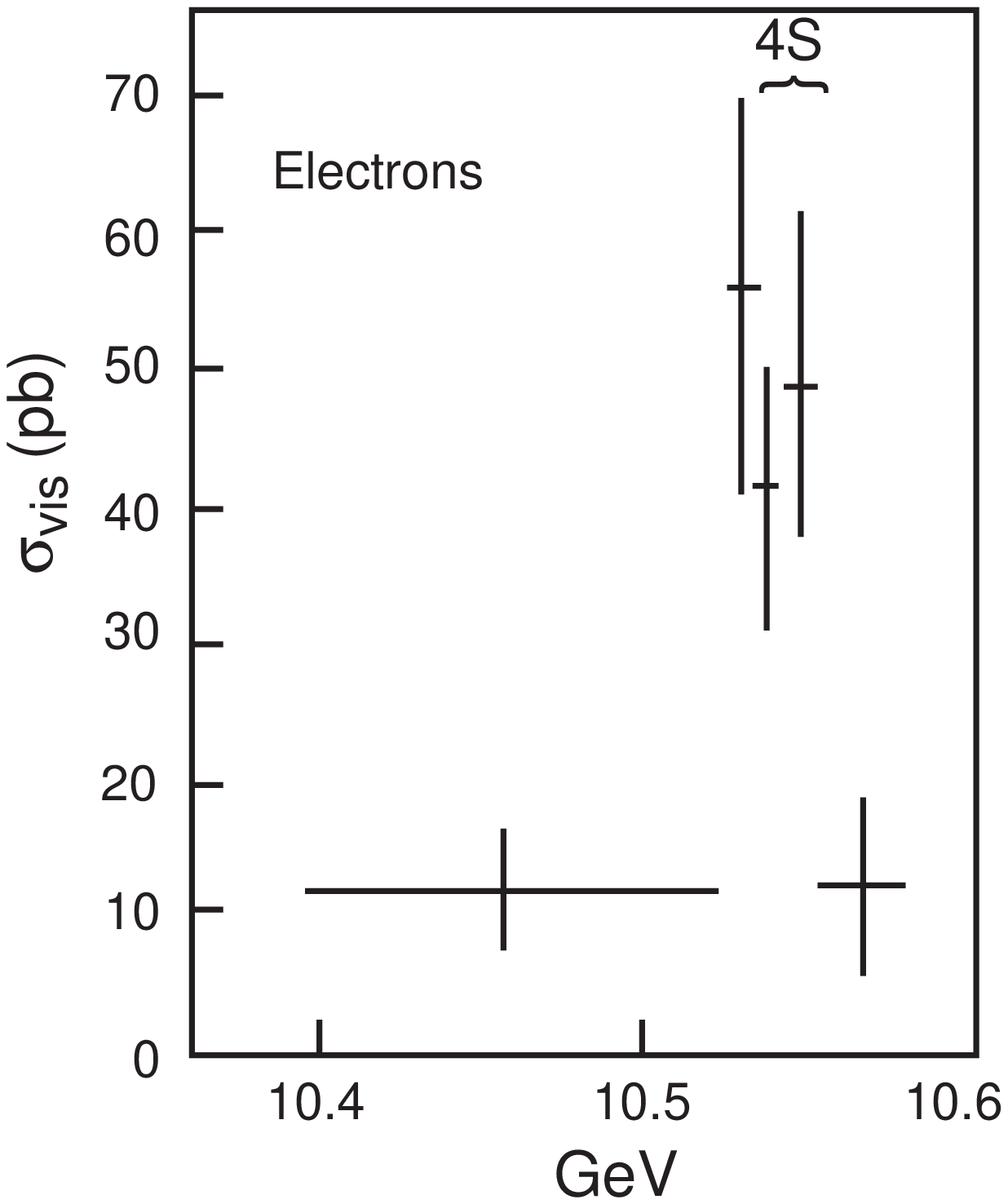, width=2in} 
\epsfig{file=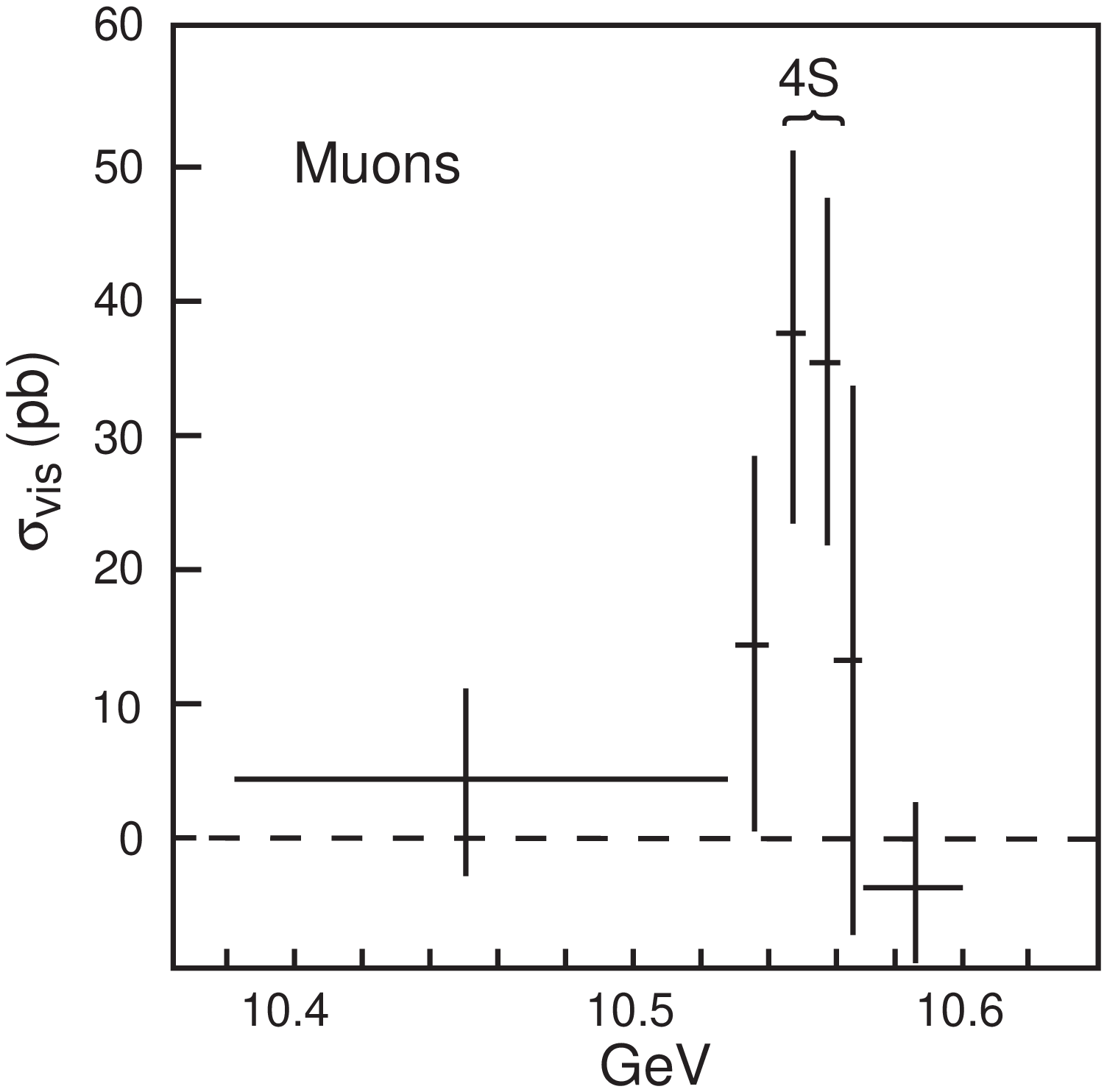, width=2in}
\end{center}
\caption{Yield of electrons (left) and muons (right) from hadronic events, as a
function of center-of-mass energy, showing enhancement at $\Upsilon$(4S)
(CLEO).}
\label{fig:4slep}
\end{figure}

\subsection{Roughing out the Qualitative Features -- 1980-88}

A series of measurements, from 1980 to 1988, determined the qualitative
features of the $b$ quark.

\medskip

\noindent {\it 2.2.1 Semileptonic Decay Branching Fraction}

\medskip

If the $b$ decays by a charged current interaction, $b \rightarrow
W_{\rm virtual} c$ or \\ $b \rightarrow W_{\rm virtual}u$, then by
simple counting of the $W_{\rm virtual}$ final states ($\overline u
d$, $\overline c s$, $\tau \nu$, $\mu \nu$, $e \nu$), allowing for a
factor of 3 for color for $\overline u d$ and $\overline c s$, one
predicts a semileptonic decay branching fraction of 1/9. Phase space
suppresses $\overline c s$ and $\tau \nu$, and hadronic final state
interactions enhance $\overline u d$ and $\overline c s$, leading to a
theoretical prediction for the semileptonic decay branching fraction
of $\approx$ 12\%. Early measurements were in qualitative
agreement. (Aside -- now, in the precision era, the measurements
appear to be 1-2\% below the theory, and that difference is not
understood.)

\medskip

\noindent {\it 2.2.2 Ruling out Topless Models}

\medskip

Giving that the bottom quark exists, is there a top quark? That was a very
real question in the early 1980's, because searches at PEP and PETRA had
come up empty, and it was (then) hard to imagine that top was more than 2-3
times heavier than bottom. Producing 
``topless models'' became an industry among theorists. Shooting them
down became an experimental responsibility.

The simplest of the topless models had $b$ a weak isospin singlet, decaying
by flavor-mixing with $s$ and $d$.  In this case, the GIM mechanism would
be inoperative, and there would be  flavor-changing neutral decays of $b$,
in particular $b \rightarrow s \ell^+ \ell^-$. Kane and Peskin derived a
lower limit on the ratio $\Gamma (b \rightarrow s \ell^+ \ell^-)/\Gamma (b
\rightarrow c \ell \nu)$, for this topless model. CLEO (1984) and Mark J
(1983) showed that the ratio was below the Kane-Peskin limit, ruling out
that model.

A more complicated topless model had $b$ a weak isospin singlet, but
decaying {\it not} by flavor mixing but by some new mechanism -- exotic
decays, which gave rise to enhanced yields of charged leptons and/or
neutrinos and/or baryons. CLEO (1983) knocked that model off, by measuring
yields of $\mu$, $e$, $p$, and missing energy.

The last stand of topless models was a particularly ugly one due to Henry
Tye. It had $b$ in a right-handed doublet with $c$. Its decays mimicked $b
\rightarrow c W^-_{\rm virtual}$ reasonably well. However, its predicted
production asymmetry, in $e^+ e^- \rightarrow 
(\gamma \ {\rm or}\ Z^0) \rightarrow b \overline b$, was very different, in
the $\gamma - Z^0$ interference region, from the predictions for $b$ a
left-handed doublet with $t$. Experiments at PETRA (1985) established the
left-handed doublet nature of $b$, killing the final topless model.
Although it wouldn't be discovered for another 10 years, by 1985 it was
clear that top {\it had} to exist.

\medskip

\noindent {\it 2.2.3 $|V_{ub}/V_{cb}|$}

\medskip

Does $b$ decay predominantly to $c (b \rightarrow c W^-_{\rm virtual})$ or
to $u (b 
\rightarrow u W^-_{\rm virtual})$?  While there was a bias favoring $b
\rightarrow c$, as of 1980 there was no strong theoretical argument
favoring $b \rightarrow c$, nor any experimental evidence.

First evidence came from the kaon yield in $B$ decay (CLEO, CUSB, 1982),
which was large, as would be expected for a $b \rightarrow c \rightarrow s$
sequence. The yield implied $|V_{ub}/V_{cb}|^2 < 0.15$.

Next evidence came from the lepton momentum spectrum. Since $u$ is lighter
than $c$, $b \rightarrow u \ell \nu$ will have a stiffer lepton spectrum
than $b \rightarrow c \ell \nu$ (see Fig. 4). By measuring the lepton
spectrum and fitting to a mix of $b \rightarrow u \ell \nu$ and $b
\rightarrow c \ell \nu$, CLEO (1984) established that $|V_{ub}/V_{cb}|^2 <
0.04$. By concentrating on the endpoint region of the spectrum, with more
data, CLEO (1987) established that $|V_{ub}/V_{cb}|^2 < 0.02$. Finally,
with still more data, CLEO (1990) saw leptons beyond the $b \rightarrow c
\ell \nu$ endpoint, establishing that $|V_{ub}/V_{cb}|^2 > 0$.

\begin{figure}[h]
\begin{center} \epsfig{file=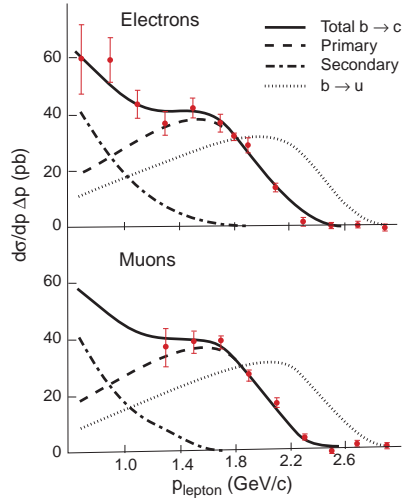, height=2.6in} \end{center}
\caption{Momentum spectrum of electrons (top) and muons (bottom) from
$\Upsilon$(4S) decays. Curves show spectra expected from $b \rightarrow u
\ell \nu$, $b \rightarrow  c \ell \nu$ (primary), $b \rightarrow c
\rightarrow s \ell \nu$ (secondary), and their sum (total $b \rightarrow c$).}
\label{fig:lep-p-spec}
\end{figure}

\medskip

\noindent {\it 2.2.4 B Reconstruction}

\medskip
  
Although there was no doubt about the existence of the $b$ quark in its
bare form, and thus no doubt about the existence of $b$-flavored hadrons,
it was nonetheless important to `reconstruct' them, to assemble the
decay products and show that they came from the decay, e.g., of a $B$ meson.
Aside from the aesthetics of ``it's got to be there, so you must {\it
show} that it is there'', $B$ reconstruction was needed to determine the $B$
meson mass. CLEO did this in 1983.

\medskip

\noindent {\it 2.2.5 b Lifetime}

\medskip

The $\Upsilon$(4S) has a mass just slightly above $B \overline B$
threshold. As a result, the {\it only} decay of $\Upsilon$(4S) is
$\Upsilon$(4S) $\rightarrow B \overline B$. There are no extra particles to
confuse the situation. However (at a symmetric $e^+ e^-$ collider) the $B$
and $\overline B$ are moving very slowly, with momenta $\sim$300 MeV/c,
$\beta \sim 0.06$. This is not a suitable environment for determining $b$
lifetime.

For $e^+ e^-$ collisions at higher energies, the $b$-flavored hadrons will
be moving faster, but the signal-to-noise will be less favorable (1 in 11,
rather than 1 in 4), and the events will be more complicated (many
particles in addition to the two $b$-flavored hadrons).  For measuring the
$b$ lifetime, the higher speed of the $b$-flavored hadrons overcomes the
disadvantages just mentioned, and makes lifetime measurements possible. In
1983, MAC and Mark II, at PEP, made such measurements. They found the
lifetime to be
{\it long}, $\approx 1$ ps, implying $|V_{cb}| \approx 0.05$. Had
$|V_{cb}|$ been more like $|V_{us}| = 0.22$, the lifetime would have been a
factor of 20 shorter, and probably would have then been too short to
measure. This result, the long $b$ lifetime, was the first surprise in $b$
physics.

\medskip

\noindent {\it 2.2.6 $B^0 - \overline B^0$ Mixing}

\medskip

It was recognized early on that the $B^0$ could, in principle, transform
itself into a $\overline B^0$, just as the $K^0$ transforms into a
$\overline K^0$. The diagram is a box diagram, with $W^+,\ W^-$ on two
parallel sides of the box, and $u,\ c,
 t,\  \overline u, \ \overline c,\ \overline t$ on the other two sides. In
the limit of equal $u,\ c,\ t$ quark masses, the summed diagrams vanish,
via the GIM mechanism. For 
``reasonable'' values of the top quark mass -- say 20 GeV, the rate for
$B^0 - \overline B^0$ mixing would be immeasurably small.

In 1987, the UA1 experiment at CERN, in $\overline p p$ collisions, saw
like-sign dilepton pairs, which they interpreted as $B^0 - \overline B^0$
mixing. As the extremely complicated environment of $\overline p p$
collisions made interpretation difficult, few people took this result
seriously.

Later that same year, the ARGUS experiment at DESY, in the clean
environment of $e^+ e^-$ collisions at the $\Upsilon$(4S), also saw
like-sign lepton pairs. {\it This} could not be ignored.  The most natural
interpretation (and the correct one) was that top was a {\it lot} heavier
than people had thought.  This interpretation was slow in being accepted.

\subsection{The Early Days -- Summary}

By 1990, there was a clear answer to ``What are the basic features of
the bottom quark?''.

\begin{itemize}

\item It is a member of a left-handed weak isospin doublet, with a ({\it
very} heavy) top quark. 

\vspace{-.1in}
\medskip

\item It decays dominantly to the charm quark, via a charged current
interaction $b \rightarrow c W^-_{\rm virtual}$. $|V_{cb}|\approx 0.04$, so
the coupling of the third generation to the second is smaller than the
coupling of the second generation to the first
$(|V_{us}| = 0.22)$.

\vspace{-.1in}
\medskip

\item Its decay to the up quark, $b \rightarrow u W^-_{\rm virtual}$, is
small but not zero, so the coupling of third generation to first is the
smallest of the three couplings.

\vspace{-.1in}
\medskip

\item Because the top quark is so massive, the GIM mechanism breaks down
for loop and box diagrams involving $b$. One consequence is the observed
large rate for $B^0 - \overline 
B^0$ mixing.  Another should be measurable rates for penguin decays. Lets
look!

\end{itemize}

\section{A Change in Objective}

In ``the early days,'' the objective was to determine the basic
features of the bottom quark. By 1990, that had been done, and the emphasis
shifted.

Now, in recent times, and in the future, the objective is to {\it use} the
bottom quark to probe the Standard Model, and search for physics `Beyond
the Standard Model'.

There are two approaches to this probing and searching.  These are:

\begin{itemize}

\item ``Overdetermining the CKM Matrix''

\vspace{-.1in}
\medskip

\item Measuring rates for Electroweak Penguins 

\end{itemize}

\noindent I'll examine each of these a bit later, but first a brief review
of the flavor sector of the Standard Model.

\subsection{The Flavor Sector of the Standard Model}

Quarks come in left-handed weak isospin doublets, and decay via emission of
(real or virtual) $W^\pm$ bosons.

$$\bigg( {t \atop b}\bigg) \  t \rightarrow bW^+\ \ ,\ \ \bigg( {c \atop
s}\bigg)\  c 
\rightarrow sW^+ \eqno(3.1\hbox{-}1)$$

\noindent Thus $t$ decays to $b$ and a real $W^+$, while $c$ decays to $s$
and a virtual $W^+$.
\medskip

\noindent {\bf Question:} How do the lower, lighter members of the doublets
decay? The $b$ quark can't decay $b \rightarrow t W^-$; that violates
energy conservation.

\medskip

\noindent {\bf Answer:} The mass eigenstates $(b,s,d)$ and the weak
interaction eigenstates $(b^\prime,s^\prime, d^\prime)$ are slightly
different. The $b$ quark ``flavor mixes'' with $s$ and $d$ quarks,
according to the CKM matrix (Cabibbo, Kobayashi, Maskawa):

\[ \left( \begin{array}{c} d^\prime \\ s^\prime \\ b^\prime \end{array}
\right)
 = \left( \begin{array}{ccc}V_{ud} & V_{us} & V_{ub} \\
                            V_{cd} & V_{cs} & V_{cb} \\
                            V_{td} & V_{ts} & V_{tb} \end{array} \right)
\left( \begin{array}{c} d \\ s \\ b \end{array} \right) \]

\hfill (3.1-2)
\vskip 0.08in
\hskip 1.3in $\Uparrow$ \hskip 1.4in $\Uparrow$

\hskip 1.1in wk. int. \hskip 1.11in mass
\medskip

\noindent So, the $b$ quark (mass eigenstate) is a mixture of $b^\prime$
(dominant component, amplitude $\propto V_{tb}$), $s^\prime$ (amplitude
$\propto V_{cb}$), and $d^\prime$ (amplitude $\propto V_{ub}$), and decays
$b \rightarrow c W^-_V$ with an amplitude proportional to $V_{cb}$, due to
its $s^\prime$ component, and decays $b \rightarrow u W^-_V$ with an
amplitude proportional to $V_{ub}$, due to its $d^\prime$ component.

The CKM matrix is a unitary matrix, which places $n^2$ constraints on an $n
\times n$ matrix.

Because the phase of each quark state is arbitrary, $2n - 1$ phases in the
CKM matrix can be transformed away.

Thus, if there were only two families of quarks, $(u,d)$, and $(c,s)$, the
CKM matrix would be a $2 \times 2$ matrix, with 4 complex elements, 8
parameters. The unitarity of the CKM matrix reduces 8 to 4, and the
arbitrariness of the quark state phases reduces 4 to 1,  a single
parameter, the Cabibbo angle. The CKM matrix for two families is described
by a single parameter, and can be made real.

But there are (at least) three families of quarks, $(u,d),\ (c,s),\ {\rm
and}\ (t,b)$. The CKM matrix is a $3 \times 3$ matrix, and the arithmetic
goes 9 complex elements $\rightarrow$ 18 parameters $-$9, for unitarity,
$-$5, for arbitrary phases $=$ 4. The CKM matrix for three families is
described by four parameters: 3 angles and 1 phase. {\it This phase cannot
be transformed away. If the phase is nonzero, weak decays will not be CP
invariant.}

Thus
$${\rm 3\ families}\ \Rightarrow \ {\rm CP\ violation.}$$

\noindent This was Kobayashi and Maskawa's insight in 1973, before the
charm quark had been discovered, let alone any members of the third family.

\medskip

\noindent {\it 3.1.1 CP Violation}

\medskip

CP violation was observed in neutral kaon decay in 1964, by Christenson,
Cronin, Fitch and Turlay. Given Kobayashi and Maskawa's insight, that
implies 3 families.

The $b$ quark, hence a third family, was observed in 1977, by Lederman and
collaborators. That implies CP violation.

So, why is everyone making such a fuss about CP violation? It's expected,
observed, explained, isn't it?  There are two reasons why CP violation is
now considered a ``big deal''.

\medskip 

\noindent A) The CP violation given by the phase in the CKM matrix appears
to be too small to account for the baryon-antibaryon asymmetry of the
Universe at early times. Cosmology requires ``New Physics,'' and it
must be CP-violating New Physics.

\medskip

\noindent B) Measurements using CP-violating  $b$ decays can help determine
(overdetermine) the CKM matrix, hence probe the correctness of the Standard
Model (or see New Physics).

\medskip

Reason B) has CP violation in $b$ decay a useful tool for probing,
searching, while reason A) has it as the primary object of study.

My own view is that too much attention is give to A) (hey, it's got a lot of
PR value), and not enough to B). But, in any case, whether you prefer to
focus on A) or B), what studies you'll perform will be much the same. CP
violation in $b$ decay should be, and will be, studied.

\medskip

\noindent {\it 3.1.2 Penguins, $B^0 - \overline B^0$ Mixing and the GIM
Mechanism}

\medskip

The GIM mechanism causes flavor-changing neutral currents to vanish at tree
level. It also suppresses FCNC beyond tree level, for loop and box diagrams.
Let's work this through for an important example, $b \rightarrow s \gamma$.
(The same argument applies to gluonic penguins $b \rightarrow sg$.)

\begin{figure}[h]
\begin{center} \epsfig{file=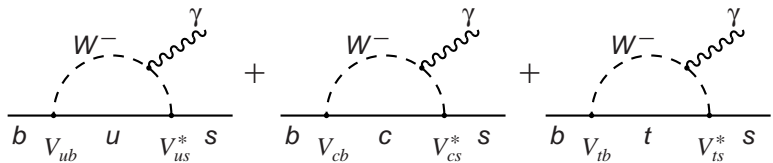, width=4in} \end{center}
\caption{Standard Model Feynman diagrams for $b \rightarrow s \gamma$
decay, with CKM factors indicated.}
\label{fig:bsgamfeyn}
\end{figure}

The Feynman diagrams for $b \rightarrow s \gamma$ are shown in Fig. 5 The
overall amplitude is the sum of the three diagrams, with $u$, $c$, and $t$
quark inside the loop. The CKM factors are as shown on the figure,
and thus the amplitude is

$$A = A \big(m^2_u \big)V_{ub} V_{us}^* + A \big( m^2_c \big) V_{cb}
V^*_{cs} + A \big( 
m^2_t \big) V_{tb} V^*_{ts} \eqno(3.1.2\hbox{-}1)$$

\noindent The amplitudes $A(m^2_u),\ A(m^2_c),\ A(m^2_t)$ depend only on
masses, their flavor dependence having been removed by factoring out the
CKM pieces. But, from unitarity

$$V_{ub} V^*_{us} + V_{cb} V^*_{cs} + V_{tb} V^*_{ts} = 0 \ \ {\rm
(Unitarity)}
\eqno(3.1.2\hbox{-}2)$$

\noindent Thus, if $m_u = m_c = m_t$, then $A=0$. That's GIM, the
cancelation of the different terms in the sum. There is suppression, and
the closer the three masses are to each other, the more the suppression.

But, $m_t \gg m_c$, $m_u$. So, the cancelation is far from complete. The
amplitude $A$ is proportional to $m^2_t$, and, since $m_t$ is a lot larger
than originally expected, penguins in $b$ decay are also a lot larger than
originally expected.

A similar argument applies to $B^0 - \overline B^0$ mixing. The Feynman
diagrams are shown in Fig. 6.  There is a double sum over $u,\ c,\ t;\
\overline u,\ \overline c,\ \overline t$.  If $m_u = m_c = m_t$, the sum is
zero, due to the unitarity of the CKM matrix. The heavy top badly breaks
GIM, with an amplitude for mixing proportional to $m_t^2$.

\begin{figure}[h]
\begin{center} \epsfig{file=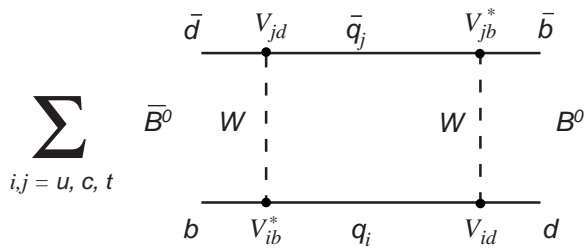, width=3in} \end{center}
\caption{Standard Model Feynman diagrams for $B^0 - \overline B^0$ mixing,
with CKM factors indicated.}
\label{fig:mixingfeyn}
\end{figure}

\subsection{Why is Bottom Quark Physics so Interesting?}

We're now in a position to answer the question ``Why is bottom quark
physics so interesting, such a good probe of New Physics?''

The answer is, ``Because the {\it TOP} quark is so massive!''

The massive top quark gives rise to substantial $\overline B^0 - \overline
B^0$ mixing, and substantial rates  from loop diagrams (Penguins). Both of
these are powerful tools for testing the Standard Model, for searching for
New Physics.

\medskip

\noindent {\it 3.2.1 Using $B^0 - \overline B^0$ Mixing to Learn Weak Phases}

\medskip

Consider a decay of a neutral $B$, with $B^0$ and $\overline B^0$ reaching
the same final state, $B^0 \rightarrow X$ and $\overline B^0 \rightarrow
X$. Examples are $B^0 \rightarrow \psi K^0 \rightarrow \psi K^0_s$ and
$\overline B^0 \rightarrow \psi 
\overline K^0 \rightarrow \psi K^0_s$; and $B^0 \rightarrow \pi^+ \pi^-$
and $\overline B^0 \rightarrow \pi^+ \pi^-$. A particle born as a $B^0$ has
two routes to this final state: i) The direct one $B^0 \rightarrow X$, and
ii) the indirect one, through $B^0 - \overline B^0$ mixing, $B^0
\rightarrow \overline B^0 \rightarrow X$.
The amplitudes for these two routes will add coherently, and interfere.
Similarly, the amplitudes for $\overline B^0 \rightarrow X$ and $\overline
B^0 \rightarrow B^0 \rightarrow X$ will add coherently and interfere.
Immediately after birth, a particle born as a $B^0$ will be a $B^0$, but,
over time, it will mix into $\overline B^0$, and so time development is the
key. By tagging particle flavor at birth, comparing $|A (B^0 \rightarrow X)
+ A(B^0 \rightarrow \overline B^0 \rightarrow X)|^2$ with $|A (\overline
B^0 \rightarrow X) + A(\overline B^0 \rightarrow B^0 \rightarrow X)|^2$,
studying the time development of both, one can determine
$$\sin \big(\phi_{\rm Mixing} + 2 \phi_{B^0 \rightarrow X}\big)$$
The expected value of $\phi_{B^0 \rightarrow \psi K_s}$ is zero, while the
Standard Model value of $\phi_{\rm Mixing}$ is $2 \beta$, so, studying $B
\rightarrow \psi K^0_s$ is the much talked about ``measurement of
$\sin 2 \beta$''.

Note that one must study {\it time development}. This class of
measurements, time development of tagged $B^0,\ \overline B^0$ to a common
final state, is the rationale behind {\it asymmetric} $e^+ e^-$ colliders
at the $\Upsilon$(4S). The asymmetric initial state energies has the
center-of-mass moving in the lab, so the decay time can be measured.

\medskip

\noindent {\it 3.2.2 Electroweak Penguins as Probes of New Physics at High
Mass Scales}

\medskip

The Standard Model diagram for 
$b \rightarrow s \gamma$ is shown in Fig. 7. The photon may be emitted from
any of the charged lines. The top quark internal line is shown, because it
is the excess of $m^2_t$ above $m^2_c,\ m^2_u$, that breaks the GIM
mechanism. The mass scale of the diagram is set by the masses of the
particles in the loop -- $m_t$ and $M_W,\ \sim 100$ GeV.

\begin{figure}[h]
\begin{center} \epsfig{file=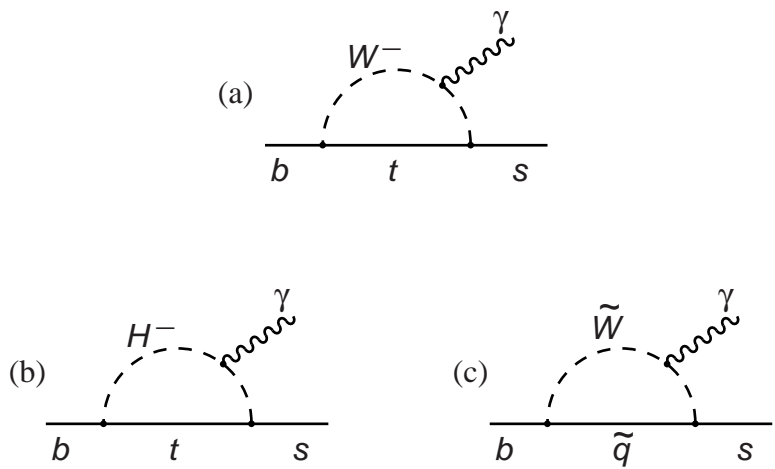, width=4in} \end{center}
\caption{(a) The dominant Standard Model Feynman diagram for $b \rightarrow
s \gamma$. (b) A possible ``New Physics'' contribution -- charged
Higgs. (c) A possible ``New Physics'' contribution -- SUSY.}
\label{fig:NewPhys}
\end{figure}

Consequently, contributions from New Physics (e.g., charged Higgs, SUSY
particles; see Fig. 7) will show up for New Physics masses in that same
range. So, penguins probe for New Physics up to masses $\sim$500 GeV.

This argument, given for electroweak penguins, applies also to gluonic
penguins.  However, electroweak penguins, in particular $b \rightarrow s
\gamma$, has the advantage that its rate can be calculated, within the
Standard Model, and Beyond the Standard Model, to a precision of 10\%.

\newpage

\noindent {\it 3.2.3 Learning Weak Phases from Penguin-Tree Interference}

\medskip

Many rare $B$ decays involve both penguin and tree amplitudes, while some
related decays are pure penguin, or pure tree. By studying relative rates
and CP asymmetries, one can sort the phases out, and determine weak phases.

As an example, consider $B^- \rightarrow K^- \pi^0$, $B^+ \rightarrow K^+
\pi^0$, and $B^\pm \rightarrow K^0 \pi^\pm$. The first two involve both
penguin and tree amplitudes, while the last is pure penguin. The amplitudes
for the three processes are 
$$A(B^-  \rightarrow K^- \pi^0) = A_P + A_T e^{i\phi_W} e^{i\phi_S}$$
$$A(B^+ \rightarrow K^+ \pi^0) = A_P + A_T e^{-i\phi_W} e^{i\phi_S}
\eqno(3.2.3\hbox{-}1)$$
$$A(B^+ \rightarrow K^0 \pi^+) = A_P$$
where $\phi_W$ is the difference in weak phase between penguin and tree,
and $\phi_S$ is the difference in strong phase between penguin and tree.
Squaring amplitudes, one sees, for the first two modes
$$\vert A \vert ^2 = A^2_P + 2 A_P A_T [\cos \phi_W \cos \phi_S \mp \sin
\phi_W
\sin \phi_S] + A^2_T \eqno(3.2.3\hbox{-}2)$$
while the square for the third mode is just $|A_P|^2$.

Thus, the rate difference, i.e., CP asymmetry, gives $\sin \phi_W \sin
\phi_S$, while the rate sum, compared with the third mode, gives $\cos
\phi_W \cos \phi_S$. Of course, there are complications to the naive
picture just presented, due to electroweak penguins, color-suppressed
trees, and long distance rescatterings.  But, the decays {\it do} depend
on strong and weak phases,
roughly as indicated, and by studying several rare decays one can learn
weak phases.

\subsection{How Does One Determine Elements of the CKM Matrix?}

$$\pmatrix{V_{ud} & V_{us} & V_{ub}\cr
                            V_{cd} & V_{cs} & V_{cb} \cr
                            V_{td} & V_{ts} & V_{tb}\cr}\eqno(3.3\hbox{-}1)$$ 

\noindent Rates for nuclear beta decay, compared to the rate for muon
decay, gives a very precise determination of the magnitude of $V_{ud}$.
Kaon and hyperon decay rates give good determinations of the magnitude of
$V_{us}$. Assorted studies of charm decay give rough measurements of
$|V_{cd}|$ and $|V_{cs}|$. However, since the third generation is only
weakly coupled to the first two, these studies determine only a single
parameter, $\lambda = \sin \theta_{\rm Cabibbo}$.

Studies of $b$ decay determine two more parameters. In particular, the rate
for $b \rightarrow c \ell \nu$ determines $|V_{cb}|$, and the rate for $b
\rightarrow u \ell \nu$ determines $|V_{ub}|$. 

Can $|V_{ts}|$, $|V_{td}|$ be determined from studies of top decay? {\it
Not soon!}  The rate for $t \rightarrow W^+ s$ is proportional to
$|V_{ts}|^2$, and the rate for $t \rightarrow W^+ d$ is proportional to
$|V_{td}|^2$. Measuring those rates would give $|V_{ts}|$ and $|V_{td}|$.
But the expected value for $|V_{ts}|^2$ is $\sim 2 \times 10^{-3}$, while
that for $|V_{td}|^2$ is $\sim 10^{-4}$,
while $|V_{tb}|^2 \approx 1$, giving a dominant decay $t \rightarrow Wb$.
 It will be a while (quite a while) before top decay branching fractions at
the $10^{-3} - 10^{-4}$ level are measured.

So, for the foreseeable future, the situation is this.  We can determine
three magnitudes in the CKM matrix -- $\lambda$, $|V_{cb}|$, $|V_{ub}|$ --
from tree-level processes, theoretically secure, relatively free from
possible New Physics contributions, reliably giving what they claim to
determine. All else will come from loops, boxes, places where New Physics
is likely to enter. Thus, if an ``overdetermination of the CKM matrix''
finds an inconsistency, that does {\it not} mean a problem with the CKM
matrix, but rather that the relation to the CKM matrix of some measurable
has been changed by New Physics. For example, if $\sin 2 \beta$ as
determined from the time development of tagged $B^0 \rightarrow \psi K^0_s$
disagrees with expectations,  that would mean that the phase of the $B^0 -
\overline B^0$ mixing amplitude is {\it not} $\sin 2 \beta$, but has been
altered by New Physics contributions to mixing.

Let's rewrite the CKM matrix in a $b$-centric fashion. Taking $|V_{cb}| = 0
(\lambda^2)$, $|V_{ub}| = 0 (\lambda^3)$, and enforcing unitarity, we have,
correct to $0 (\lambda^3)$
$$\pmatrix{1-\lambda^2/2 & \lambda &
\vert V_{ub} \vert e^{-i\gamma}\cr
                            - \lambda & 1 - \lambda^2/2  & \vert V_{cb}
\vert \cr
\lambda \vert V_{cb} \vert  - \vert V_{ub} \vert e^{i \gamma}&
- \vert V_{cb} \vert & 1\cr}\eqno(3.3\hbox{-}2)$$
Since $\lambda$ is already determined with high precision, this form makes
apparent the urgency of good determinations of $|V_{cb}|$ and $|V_{ub}|$.

$|V_{cb}|$ is obtained from measurements of the $B$ meson lifetime, and
either the rate for $B \rightarrow D^* \ell \nu$ extrapolated to the point
where $D^*$ is at rest, or the rate for $B \rightarrow X \ell \nu$
inclusive, plus information on the $b$ quark mass and HQET Operator Product
Expansion parameter $\lambda_1$.  The $b$ lifetime and semileptonic decay
branching fraction are well measured. CLEO has in hand data on $B
\rightarrow D^* \ell \nu$, and on moments of hadronic mass and lepton
energy in $B \rightarrow X \ell \nu$ sufficient for $\pm 4$\%
determinations of $|V_{cb}|$, separately by each method. For now, $|V_{cb}|
= (39.5 \pm 3.6) \times 10^{-3}$.

$|V_{ub}|$ is less well determined, and $\gamma$ even less well determined.
\newpage

\section{What is {\boldmath $|V_{ub}/V_{cb}|$}?}

\subsection{Limitations of Previous Approaches}

In Section 2.2.3, I described progress during the early days in placing
upper-limits on, and finally establishing a nonzero value for,
$|V_{ub}/V_{cb}|$. All the approaches tried then had serious limitations.
The kaon yield approach was really a measurement of $|V_{cb}|$, limiting
$|V_{ub}|$ by $|V_{cb}|$'s deviation from 1.0.  Since the total number of
kaons produced per $b \rightarrow c W_V$ decay is uncertain at greater than
the ten percent level, this approach was quickly discarded, once it was
realized that $|V_{ub}/V_{cb}|$ was in the sub-ten-percent range.

Fitting the measured lepton spectrum in $B$ semileptonic decay to the
predicted spectra for $b \rightarrow c \ell \nu$ and $b \rightarrow u \ell
\nu$ hits its limit because, with the $b \rightarrow u \ell \nu$ rate less
than 5\% of the $b \rightarrow c \ell \nu$ rate, minor errors in modeling
of the $b \rightarrow c \ell \nu$ spectrum cause major errors in the $b
\rightarrow u \ell \nu$ yield. This approach has also been discarded.

The endpoint approach avoids sensitivity to the $b \rightarrow c \ell \nu$
modeling because it limits the focus to the lepton momentum range where $b
\rightarrow c \ell \nu$ is small or zero. But here there is sensitivity to
the modeling of $b \rightarrow u \ell \nu$. The fraction of the $b
\rightarrow u \ell \nu$ spectrum in the endpoint windows cannot be reliably
calculated, and its uncertainty limits accuracy of $|V_{ub}|$ by this
method to $\sim$20\%. While results from this approach are currently one of
the two ways now giving useful results, future improvements to the 10\%
range and below seem unlikely.  (Note added in proof:  Leibovich, Low, and
Rothstein, hep-ph/9909404 v2, show how to determine the fraction, using a
measurement of the photon spectrum from $b \rightarrow s \gamma$.)

\subsection{Neutrino Detection}

The difficulty in studying $b \rightarrow u \ell \nu$ is the neutrino. If
that particle were `detected', its momentum measured, then the decay
would cause no problems. Consequently, several of us in CLEO are attempting
a new approach, `detecting' the neutrino in a semileptonic decay via the
missing 4-momentum in the event. Given a 
`detected' neutrino, one can then carry out full reconstruction of
exclusive semileptonic decays, or look at the mass distribution in
inclusive semileptonic decays.

\medskip

\noindent {\it 4.2.1 Exclusive Decays $B \rightarrow \pi \ell \nu$, $B
\rightarrow \rho \ell \nu$}

\medskip

With the neutrino `detected', the decays $B \rightarrow \pi \ell \nu$,
$B \rightarrow \rho \ell \nu$, $B \rightarrow \omega \ell \nu$ have no
undetected particles, and so the standard $B$ reconstruction technique is
applicable. The summed energy of the decay products of the candidate $B$
are compared with the beam energy, giving a difference 
$\Delta E$ which should peak at zero.  The summed vector momenta of the
decay products of the candidate $B$, {\boldmath $P_{\rm cand}$}, are used
to calculate the ``beam constrained mass'' $M_{m \ell \nu} =
\sqrt{E^2_{\rm beam} - P^2_{\rm cand}}$, which should peak at the $B$ mass.

We completed and published an analysis for $B \rightarrow \pi \ell \nu$,
and $B \rightarrow \rho/\omega \ell \nu$ some time ago (PRL 77, 5000 (16
Dec. 1996)), based on a 4$fb^{-1}$ data sample. The plots of mass and
energy difference are shown in Fig. 8.  The branching fraction accuracy
(statistical plus systematic) gave a 12\% uncertainty in $V_{ub}$, and that
uncertainty should fall as $1/\sqrt{{\cal L}}$. 

\begin{figure}[h]
\begin{center} \epsfig{file=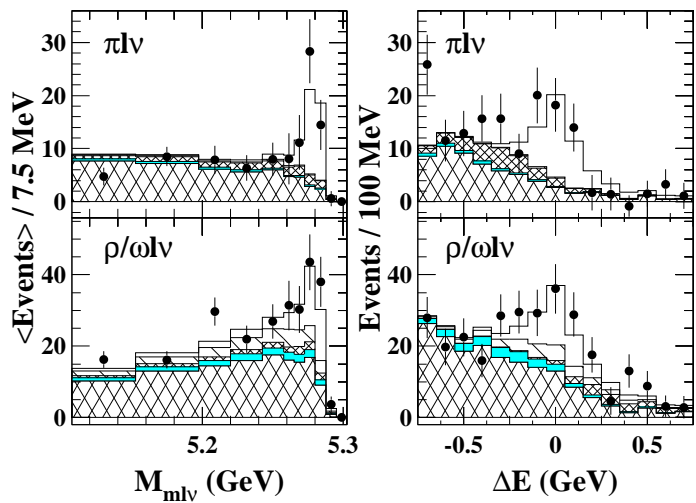, width=4.5in} \end{center}
\caption{$B$ mass and energy difference plots for $B \rightarrow \pi \ell
\nu$ and $B \rightarrow \rho \ell \nu$ (CLEO).}
\label{fig:pirhoelnu}
\end{figure}

The big issue is model dependence -- how the branching  fraction for the
exclusive modes are related to $V_{ub}$. Of course, they are proportional
to $|V_{ub}|^2$, with the constants of proportionality related to form
factors. It is through the uncertainty in the form factors that model
dependence enters.  For the 1996 analysis, we estimated this at $\pm$20\%
in $|V_{ub}|$.  This will improve with more data, which will allow
measurement of the $q^2$ dependence of the decays, providing constraints on
models for form factors. It will also improve with better form factor
calculations, from lattice gauge QCD and other techniques. Finally, studies
of the decays $D \rightarrow \pi \ell \nu$, $D \rightarrow \rho \ell \nu$,
where $|V_{cd}|$ is quite well known, can also help.  One can expect an
accuracy in $|V_{ub}|$ from CLEO's existing, 14$fb^{-1}$ data sample, in
the $\pm$15\% range, or better, depending on how much progress can be made
on the model dependence.

\medskip

\noindent {\it 4.2.2 Inclusive Decays, $B \rightarrow X_u \ell \nu$}

\medskip

Given a `detected' neutrino, and a (really) detected charged lepton, one
can calculate the mass of the hadronic system $X$ in the decay $B
\rightarrow X \ell \nu$:
$$M^2_X = M^2_B + M^2_{\ell \nu} - 2 E_B E_{\ell \nu} + 2 P_B P_{\ell \nu}
\cos
\theta_{\ell \nu, B}$$
All quantities in this equation are known except $\theta_{\ell \nu , B}$,
the angle between the $B$ meson and the $\ell \nu$ system (everything
evaluated in the lab frame). The total lack of knowledge of $\theta_{\ell
\nu , B}$ results in a smearing in the determination of $M^2_X$, which is
reasonably small since $P_B$ is small ($\sim$300 MeV/c).

The game plan, then, is to measure the $M^2_X$ distribution, given neutrino
and charged lepton, and then fit that distribution with a sum of $b
\rightarrow  u \ell \nu$ and $b \rightarrow c \ell \nu$. The contributions
from $b \rightarrow c \ell \nu$ will include $D,\ D^*,$ and heavier stuff.
The contributions from $b \rightarrow u \ell \nu$ will dominantly be below
the $D$ meson mass, consisting of $X_u$ objects like $\pi,\ \rho,\ A_1,\
A_2,$ etc. A calculation of the expected $X_u$ mass distribution is
possible, for example from a naive spectator model, or more properly from
HQET and OPE. If one could measure $M_X$ to high precision, separating $b
\rightarrow u \ell \nu$ from $b \rightarrow c \ell \nu$ would be easy, and
an inclusive measurement of $|V_{ub}|$, with relatively little model
dependence, would be possible.

Unfortunately, the measurement of $M^2_X$ so far achieved has rather poor
resolution, due to the inaccuracy in determining the neutrino vector
momentum. This inaccuracy is not so much from the inaccuracy in measurement
of individual particles, but rather from `messups' (inefficiencies in
detecting charged particles and photons, false tracks and photons), and
also from undetectable particles ($K$-long, neutrons, second neutrinos in
the event). The consequence of the poor resolution in $M_X$ is that there
is a low-mass-side tail to $b \rightarrow c \ell \nu$, which swamps the $b
\rightarrow u \ell \nu$ contribution.

An analysis has been completed (Scott Roberts' Ph.D. thesis, University of
Rochester, 1997), but not submitted for journal publication.  To suppress
the $b \rightarrow c \ell \nu$ component, we required $P_\ell > 2.0$ GeV/c,
a momentum bite a factor of 2 bigger 
than the $P_\ell > 2.3$ GeV/c typical of an endpoint analysis. The choice
of 2.0 GeV/c was a compromise between reducing model dependence (wanting a
lower momentum cut) and suppressing $b \rightarrow c \ell \nu$ (wanting a
higher momentum cut).

\begin{figure}[h]
\begin{center} \epsfig{file=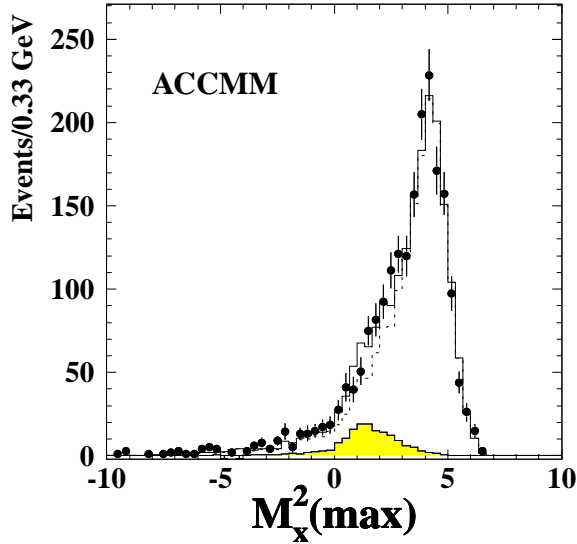, width=3in} \end{center}
\caption{$M^2_X$ distribution from decay $B \rightarrow X \ell \nu$. The
curves are $b \rightarrow u \ell \nu$ (shaded) and $b \rightarrow c \ell \nu$
(dashed curve) contributions, and their sum (Scott Roberts' thesis).}
\label{fig:MXSQ-Scott}
\end{figure}

The measured $M^2_X$ distribution is shown in Fig. 9.  The fitted
components from $b \rightarrow c \ell \nu$ and $b \rightarrow u \ell \nu$
are also shown. In the region where $b \rightarrow u \ell \nu$ is
substantial, the $b \rightarrow c \ell \nu$ background is about twice the
$b \rightarrow u \ell \nu$ signal.  Taking faith in our modeling of the $b
\rightarrow c \ell \nu$ background (though allowing a systematic error for
its uncertainty), we obtained a fit, from a 5$fb^{-1}$ data sample, which
gives $|V_{ub}|$ to $\pm$16\% -- statistical plus systematic. We did not
carry out a careful study of model dependence, but since the lepton
momentum bite is twice as large as that for the endpoint analysis, one
would expect a model dependence that is twice as small -- $\pm$10\% instead
of $\pm$20\%.

The value of $|V_{ub}|$ obtained from the fit is quite reasonable, and the
combined nominal error, ($\pm$16\% with $\pm$10\%) are competitive. But the
plot is certainly not very convincing. The $b \rightarrow c \ell \nu$
component is just too large in the $b \rightarrow u \ell \nu$ signal
region. And we would like to push the lepton momentum cut down, say to 1.8
or 1.6 GeV/c, which would make the $b \rightarrow c \ell \nu$ background
several times larger. So, our plan is not to publish this analysis, but to
work on it some more -- a {\it lot} more.

\begin{itemize}

\item We will use the full CLEO II data sample of 14$fb^{-1}$, a factor of
3 increase from that in Scott Roberts' analysis. (This is the easy one.)

\vspace{-.1in}
\medskip

\item We will improve the accuracy with which neutrinos are `detected'
and their momenta determined, by upgrading our algorithm for distinguishing
between showers in the electromagnetic calorimeter caused by photons and by
hadrons; by improving various aspects of charged particle tracking; and by
pushing to lower momentum our electron identification capabilities (we veto
events with more than one charged lepton, hence more than one neutrino).

\vspace{-.1in}
\medskip

\item Finally, we will study the correctness of our simulation of the $b
\rightarrow c \ell \nu$ component, to be sure we are correctly modeling
the low-mass tail. (For example, we will fake $K$-long events by finding
events with $K$-shorts, then pitching the $K$-short, and see if the $M^2_X$
spectra so obtained for data and Monte Carlo agree.)

\end{itemize}

The original motivation for neutrino `detection' was for studying
inclusive decays, with its use for exclusive decays an afterthought. We
still view the inclusive approach as the best hope for a measurement of
$|V_{ub}|$ to $\pm$10\% accuracy.

\section{Rare Hadronic {\boldmath $B$} Decays}

\subsection{Introduction}

I should start this section by saying what I mean, and indeed what is
typically meant, by ``rare'', as it refers to $B$ decays. A \lq\lq
rare'' $B$ decay is one which involves penguin or box diagrams. With this
definition, it is easy to see why the field of rare $B$ decays is ahead of
the field of rare kaon decays, why $b \rightarrow s$ processes have been
studied, while $s \rightarrow d$ processes much less so. The CKM factor for
$b \rightarrow s$ penguins is $V_{tb} V^*_{ts}$, while that for the
dominant, $b \rightarrow c$ is $V_{cb}$. $|V_{tb}V_{ts}^*/V_{cb}|^2 \sim
1$. For kaon decays, the penguin with top quark in the loop has a CKM
factor $V_{ts} V^*_{td}$, while that for the dominant, $s \rightarrow u$
tree is $V_{us}$. $|V_{ts} V^*_{td}/V_{us}|^2 \sim 3 \times 10^{-6}$, so
the branching fractions for rare $B$ decays are typically 5-6 orders of
magnitude larger than those for rare kaon decays -- $10^{-5} - 10^{-6}$ vs
$10^{-11} - 10^{-12}$.

As we saw in Section 3.2.3, rare decays involving penguins often also
involve $b \rightarrow  u$ trees (see Fig. 10). The example given there was
$B \rightarrow  K \pi$, a ``Cabibbo-allowed penguin'', i.e., a $b
\rightarrow  s$ penguin.  The tree diagram there is the \lq\lq
Cabibbo-suppressed $b \rightarrow  u$ tree'', $b \rightarrow  u W^-_V$,
$W^-_V \rightarrow  \overline u s$. The Cabibbo-suppression is in the decay
of the virtual $W$, $W^- \rightarrow  \overline u s$, rather than the
Cabibbo-allowed decay $W^- \rightarrow  \overline u d$. This same mix of
Cabibbo-allowed penguin plus (sometimes) Cabibbo-suppressed $b \rightarrow
u$ tree occurs for $B \rightarrow  K^* \pi$,
$B \rightarrow K\rho$, $B \rightarrow K \omega$, $B \rightarrow K^* \rho$.
The Cabibbo-allowed penguin diagram contributes to all of these, while the
Cabibbo-suppressed $b \rightarrow  u$ tree is absent for the charge modes
involving neutral $K$ or $K^*$.

\begin{figure}[h]
\begin{center} \epsfig{file=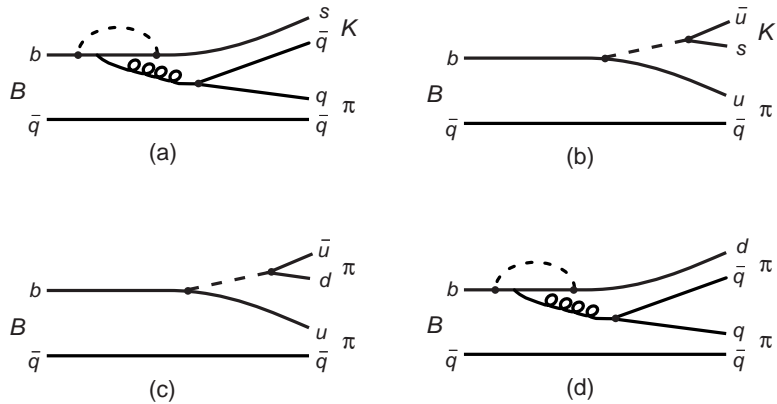, width=4in} \end{center}
\caption{The Penguin and $b \rightarrow u$ Tree diagrams contributing to
rare hadronic $B$ decays. (a) Cabibbo-allowed penguin; (b)
Cabibbo-suppressed $b \rightarrow u$ tree; (c) Cabibbo-allowed $b
\rightarrow u$ tree; (d) Cabibbo-suppressed penguin.}
\label{fig:Peng-Tree}
\end{figure}

There is also a class of decays involving Cabibbo-allowed $b \rightarrow
u$ trees, i.e., $b \rightarrow  u W^-_V,\ W^-_V \rightarrow  \overline u
d$, and Cabibbo-suppressed penguins, i.e. $b \rightarrow  d$ penguins (Fig.
10c,d). Examples include $B \rightarrow  \pi \pi$, $B \rightarrow  \pi
\rho$, $B \rightarrow  \pi \omega$, $B \rightarrow  \rho \rho$.  In fashion
similar to the `allowed penguin, suppressed tree' class, there are
particular modes for which the Cabibbo-suppressed penguin is absent,
e.g. $B^\pm \rightarrow \pi^\pm \pi^0$.

So, Penguin-Tree interference is the rule rather than the exception in rare
hadronic $B$ decays. And the exceptions, modes which are pure
allowed penguins, or pure allowed $b \rightarrow  u$ trees, help to sort
out the interference.

As mentioned in Section 3.2.3, the simple picture is complicated by
electroweak penguins $b \rightarrow  s Z^0$ (we've been talking about
gluonic penguins $b\rightarrow  sg$), color-suppressed trees, long distance
rescattering.  It will require
careful study of many rare decays before a precise value of the CKM phase
$\gamma$ can be obtained. But as we will see, some qualitative information
can already be obtained.

\subsection{The Data Sample}

CLEO has 10 million events of the form $e^+ e^- \rightarrow \Upsilon{\rm
(4S)}\ \rightarrow  B + \overline B$, and has recently completed analysis
of several of the rare decay modes. The reconstruction is conventional,
with the summed energy of the decay products of the candidate $B$ compared
with the beam energy, and the summed vector momenta of the decay products
of the candidate $B$ used to calculate the beam-constrained mass.

There are substantial backgrounds to rare $B$ decays, not from the dominant
$b \rightarrow  c$ tree decays, but from the continuum background process
$e^+ e^- \rightarrow  q \overline q$, $q = u,d,c,s$. These backgrounds are
2-jet-like, and are suppressed by a maximum likelihood fit, inputting many
`shape variables'.

Examples of $B$ mass plots and $\Delta E$ plots are shown in Fig. 11.

\begin{figure}[h]
\begin{center} \epsfig{file=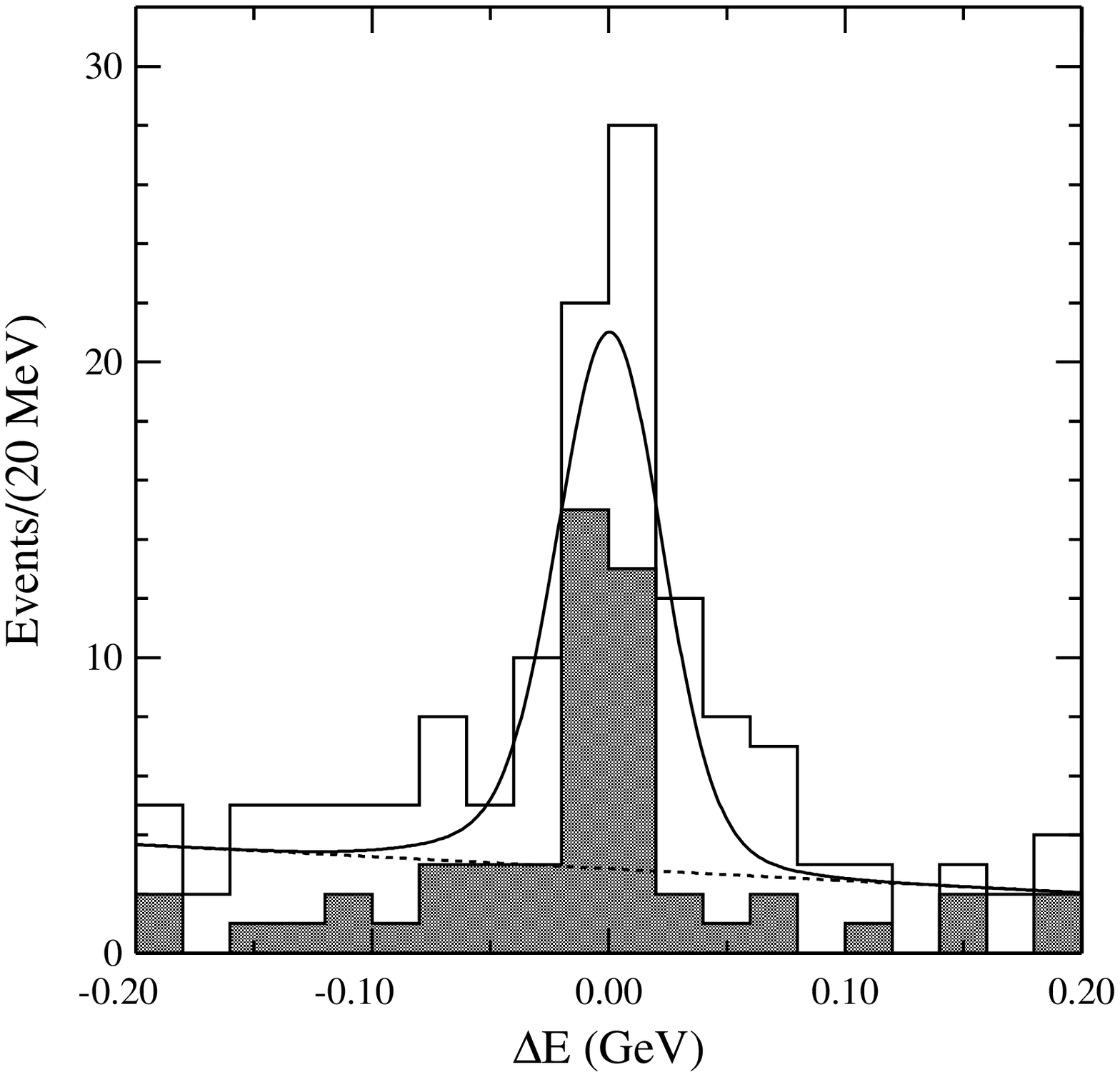, width=2in}
\epsfig{file=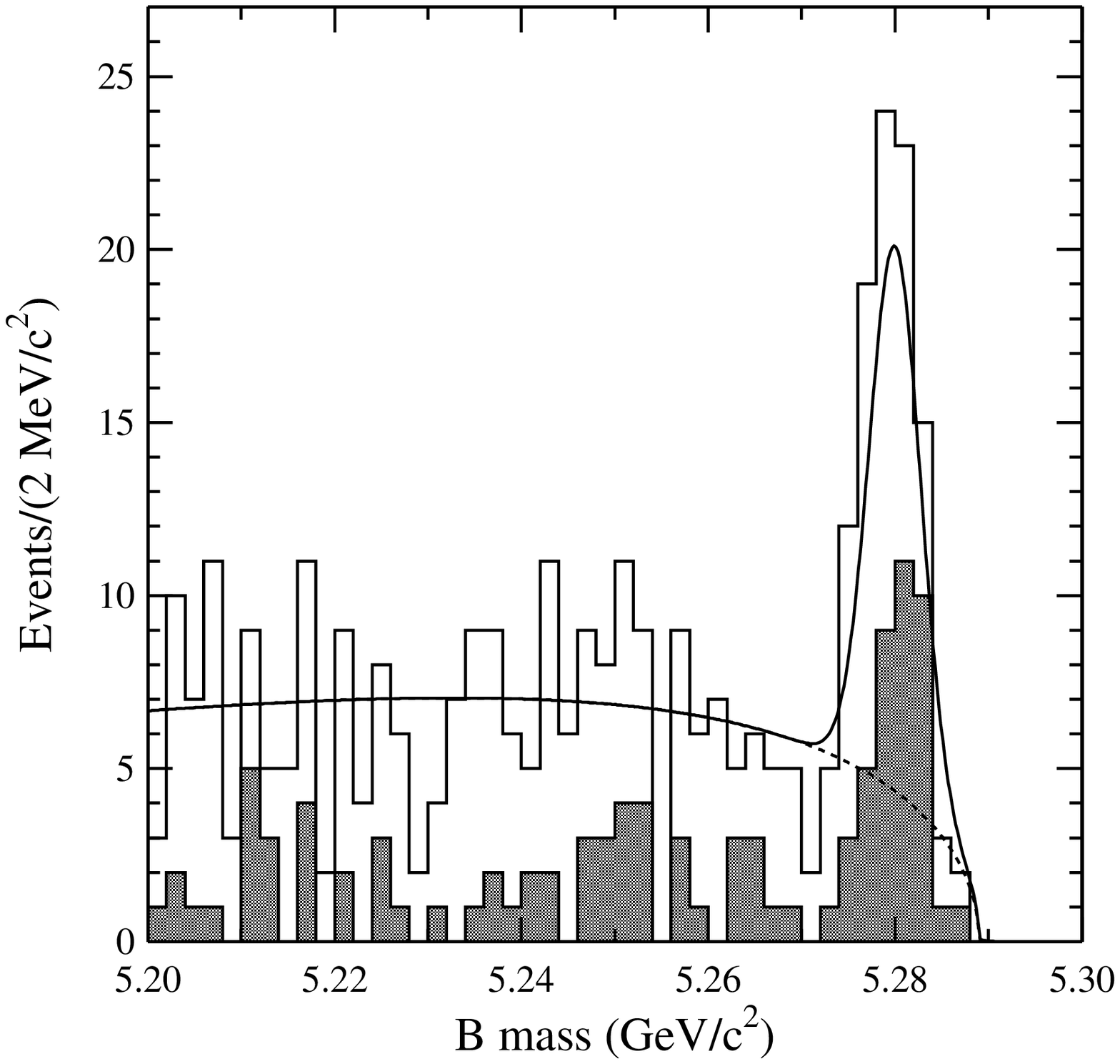, width=2in} \end{center}
\caption{Plots of (left) $\Delta E \equiv E_{\rm cand} - E_{\rm beam}$, and
(right) $M_{\rm cand} \equiv \sqrt{E^2_{\rm beam} - P^2_{\rm cand}}$, for the
decay $B \rightarrow \eta^\prime K^\pm$ (CLEO). The filled histograms
are from $\eta^\prime \rightarrow \eta \pi^+ \pi^-$, $\eta \rightarrow 
\gamma \gamma$, while the unfilled are from $\eta^\prime \rightarrow
\rho^0 \gamma$.}
\label{fig:Mb-DeltaE}
\end{figure}

\subsection{Results}

\medskip

\noindent {\it 5.3.1 $B \rightarrow  K \pi,\ \pi \pi,\ KK$}

\medskip

Results for these modes are given in Table 1 All four $B \rightarrow  K
\pi$ modes have been convincingly seen. Only one of the three $B
\rightarrow  \pi \pi$ modes has been convincingly seen, though the evidence
for $B \rightarrow  \pi^\pm \pi^0$ is fairly good. No $B \rightarrow  KK$
mode has been seen, nor were they expected to be.

\begin{table}[h]
\centering
\begin{center}
\caption{Branching fractions for $B$ decays to $K \pi$, $\pi \pi$, and $K K$.}
\end{center}
\label{tab:kpi-pipi}
\vskip 0.3cm
\begin{tabular}{lclc} \hline \hline
Mode  &  ${\cal B} (10^{-6})$ & Mode  &  ${\cal B} (10^{-6})$ \\ \hline
$K^+ \pi^-$     &   17 $\pm$ 3  &   $K^0 \pi^+$     &   18 $\pm$ 5  \\
$K^+ \pi^0$     &   12 $\pm$ 3  &   $K^0 \pi^0$     &   15 $\pm$ 6  \\
                &               &                   &               \\
$\pi^+ \pi^-$   & 4.3 $\pm$ 1.6 &  $\pi^+ \pi^0$    & 5.6 $\pm$ 3.0 \\
                &               &  $\pi^0 \pi^0$    &   $<$ 9.3     \\
                &               &                   &               \\
$K^+ K^-$       &   $<$ 1.9     &  $K^+ \bar K^0$   & $<$ 5.1 \\ \hline \hline
\end{tabular}
\end{table}

\medskip
\vfill\eject

\noindent{\it 5.3.2 $B$ Decays Involving $\eta$ or $\eta^\prime$}

\medskip

Results for the decay $B \rightarrow  (\eta\ {\rm or}\ \eta^\prime)$ ($K$
or $K^*$)  are shown in Table 2. One sees that $B \rightarrow  \eta^\prime
K$ is {\it big}, much larger than all the others.  $B \rightarrow  \eta
K^*$ is seen, and is larger than $B \rightarrow  \eta K$.

\begin{table}[h]
\centering
\caption{Branching fractions for $B$ decays to ($\eta$ or $\eta^\prime$)
plus ($K$ or $K^*$).}
\label{tab:eta-K}
\vskip 0.3cm
\begin{tabular}{lclc} \hline \hline
Mode  &  ${\cal B} (10^{-6})$ & Mode  &  ${\cal B} (10^{-6})$ \\ \hline
$B^\pm \rightarrow \eta^\prime K^\pm$     &   80 $\pm$ 12  
&   $B^\pm \rightarrow \eta^\prime K^{* \pm}$     &   $<$ 87  \\
$B^0 \rightarrow \eta^\prime K^0$     &   88 $\pm$ 19 
&   $B^0 \rightarrow \eta^\prime K^{* 0}$     &   $<$ 20 \\
$B \rightarrow \eta^\prime K$     &   83 $\pm$ 11  
&   $B \rightarrow \eta^\prime K^*$     &   $<$ 22  \\
                &               &                   &               \\
$B^\pm \rightarrow \eta K^\pm$     &   $<$ 7.1  
&   $B^\pm \rightarrow \eta K^{* \pm}$     &  27 $\pm$ 10  \\
$B^0 \rightarrow \eta K^0$     &   $<$ 9.5  
&   $B^0 \rightarrow \eta K^{* 0}$     &   14 $\pm$ 5  \\
$B \rightarrow \eta K$     &   $<$ 5.2  
&   $B \rightarrow \eta K^*$     &   18 $\pm$ 5  \\ \hline \hline
\end{tabular}
\end{table}

The interpretation of these results is far from clear.

\begin{itemize}

\item The $\eta^\prime$ could perfectly well contain a $c \overline c$
component, and if it did, a Cabibbo-allowed $b \rightarrow  c$ tree could
contribute (as it does for $B \rightarrow  \psi K)$. This situation would
lead to enhanced branching fractions for both $B \rightarrow  \eta^\prime
K$ and $B \rightarrow  \eta^\prime K^*$.

\vspace{-.1in}
\medskip

\item As pointed out by Lipkin, there will be contributions from the
gluonic penguins $b \rightarrow  sg,\ g \rightarrow  s \overline s$ and $b
\rightarrow  sg,\ g \rightarrow  u \overline u / d \overline d$, and these
diagrams will interfere. Lipkin argues that this will enhance $B
\rightarrow  \eta^\prime K$ and $B \rightarrow  \eta K^*$ relative to $B
\rightarrow  \eta^\prime K^*$ and $B \rightarrow  \eta K$.

\end{itemize} 

\noindent The data show some features of both suggestions, but at
present there is no quantitative understanding.

\medskip

\noindent{\it 5.3.3 Decays $B \rightarrow$ Pseudoscalar Vector}

\medskip

Only a smattering of these have been seen so far, e.g. $B^\pm \rightarrow
\omega \pi^\pm$, $B \rightarrow  \rho \pi$, $B \rightarrow  K^* \pi$.
CLEO's analyses of the Pseudoscalar-Vector modes  are  finished for the
full data sample of 10 million $B \overline B$ events for about half of the
decay modes. Results are given in Table 3.

\begin{table}[h]
\centering
\caption{Branching fractions for $B$ decays to Pseudoscalar Vector.
(Numbers in parentheses are preliminary, or don't satisfy CLEO's
4 $\sigma$ requirement for claiming a signal.)}
\label{tab:B_to_PV}
\vskip 0.3cm
\begin{tabular}{lclc} \hline \hline
Mode  &  ${\cal B} (10^{-6})$ & Mode  &  ${\cal B} (10^{-6})$ \\ \hline
$\pi^\pm \rho^0$     &   10.4 $\pm$ 4.0  
&   $K^\pm \rho^0$     &   $<$ 17.3  \\
$\pi^0 \rho^0$     &   $<$ 5.5  
&   $K^0 \rho^0$     &    ---  \\
$\pi^\pm \rho^\mp$     &   27.6 $\pm$ 8.9  
&   $K^\pm  \rho^\mp$     &   $<$ 32.3  \\
$\pi^0 \rho^\mp$     &   $<$ 42.6  
&   $K^0 \rho^+$     &   ---  \\
                 &      &    &         \\
$\pi^\pm \omega$     &   11.3 $\pm$ 3.4 
&   $K^\pm \omega$     &  (3.2 $\pm$ 2.3),  $<$ 7.9 \\
$\pi^0 \omega$     &   $<$ 5.8 
&   $K^0 \omega$     &  (10 $\pm$ 5),  $<$ 21 \\
            &               &               &    \\
$\pi^+ K^{* 0}$     &   $<$ 15.9 
&   $K^- K^{*0}$     &    $<$ 5.3 \\
$\pi^0 K^{* 0}$     &   $<$ 3.6 
&   $\overline K^0 K^{*0}$     &  --- \\
$\pi^+ K^{* -}$     &  (22 $\pm$ 9)
&   $K^+ K^{* -}$     &  --- \\
$\pi^0 K^{* -}$     &   $<$ 31.0 
&   $K^0 K^{* -} $     &  --- \\ \hline \hline
\end{tabular}
\end{table}

\subsection{Search for Direct CP Violation in  $B$ Decay}

If some $B$ decay has contributions from two (or more) amplitudes $A_1$,
$A_2$, with relative weak phase $\phi_W$, and relative strong phase
$\phi_S$, i.e., a total decay amplitude $A = |A_1| + |A_2| e^{i \phi_S} e^{i
\phi_W}$, then there will be direct CP violation in the decay, which will
show up as a charge asymmetry
$${\cal A} = {\Gamma (\overline B \rightarrow  X) - \Gamma (B \rightarrow
\overline X) \over \Gamma ( \overline B \rightarrow  X) + \Gamma (B
\rightarrow  \overline X)} \eqno(5.4\hbox{-}1)$$

If $|A_2| \ll |A_1|$, then
$${\cal A} \approx 2 {|A_2| \over |A_1|} \sin \phi_S \sin \phi_W
\eqno(5.4\hbox{-}2)$$
For penguin-tree interference, one expects $|A_2/A_1| \approx 0.2$. It's
less clear what to expect for $\sin \phi_S$, but in the absence of some
enhancement due to long range rescattering it will be small, probably less
than 0.25. So we expect ${\cal A}$ less than 0.1. CLEO results, for five
decay modes that have been convincingly seen, and are self-tagging, are
given in Table 4. There are no surprises. All asymmetries are consistent
with zero. There is not yet sufficient sensitivity to see CP
violations at the level expected. Since the errors are dominantly statistical,
and are based on 10 million $B \overline B$ pairs, it will likely be a while
before nonzero asymmetries are established.

\begin{table}[h]
\centering
\caption{CP Asymmetries for five rare decay modes.}
\label{tab:CP-asymm}
\vskip 0.3cm
\begin{tabular}{lc} \hline \hline
Mode  &  ${\cal A}$  \\ \hline
$B \rightarrow K^\mp \pi^\pm$     &    $-0.04  \pm  0.16$ \\  
$B \rightarrow  K^\mp \pi^0$     &    $-0.29 \pm 0.23$  \\
$B \rightarrow K^0_s \pi^\mp$     &    $+ 0.18 \pm 0.24$   \\
$B \rightarrow  K^\mp \eta^\prime$     &   $+ 0.03 \pm 0.12$   \\
$B \rightarrow \omega \pi^\mp$     &   $-0.34 \pm 0.25$   \\ \hline \hline
\end{tabular}
\end{table}

\subsection{Interpretation of Rare Hadronic $B$ Decays}

Now that CLEO has roughed out the rare $B$ decay terrain, what does it all
mean? Recall that in Section 3.2.3, the motivation for studying rare $B$
decays was given as using them to determine weak phases. What can the
existing rare $B$ decay data tell us about weak phases, in particular,
about $\gamma$, the phase of $V_{ub}^*$?

CLEO's visiting theorist George Hou and CLEO members Jim Smith and Frank
W\"urthwein have addressed that question. They assume naive factorization,
use effective-theory matrix elements, and ignore annihilation type
diagrams. With these assumptions, (and some more, mentioned below) they are
able to express the amplitudes for all two-body rare $B$ decays in terms of
a relatively small number of parameters.

The quark-level process $b \rightarrow  q_1 \overline q_2 q_3$ is
described, in effective theory, by ten parameters $a_1 \dots a_{10}$. These
are calculable within a QCD framework, and Hou et al. take two sets of
values from the literature.

The binding of $q_1 \overline q_2$ into  mesons is described by  decay
constants $f$,
$(f_\pi ,\  f_K$, $f_{K^*}$, $f_\rho,\  f_\omega,\  f_\phi)$ which   are
known. 

The binding of $q_3$ and the spectator antiquark into a meson is described
by form factors. For $B \rightarrow  PP$, there is a single form factor
$F_0^{BP}$ (but $P = \pi , K$), while for $B \rightarrow  PV$ there are two
more, $F_1^{BP},\ P = \pi ,\ K$, and $A_0^{BV},\ V = \rho,\ \omega$. Hou et
al. lean on SU(3), with breaking, to relate $F_0^{BK}$ to $F_0^{B \pi}$,
and $A_0^{B \omega}$ to $A_0^{B \rho}$.  They also use the relation
$F_0^{BP} = F_1^{BP}$, valid at $q^2 =0$. They thus describe the decays of
interest with just two form factors $F_0^{B \pi}$ and $A_0^{B \rho}$,
rather than six.

 Two of the penguin terms $(a_6,\ a_8)$ depend on the quark mass $(m_s\
{\rm or}\ m_d)$ and Hou et al. allow a free parameter $R_{su}$ to describe
this dependence.

Hou et al. thus use five free parameters: $F_0^{B \pi},\ A_0^{B \rho},\
R_{su},\ |V_{ub}/V_{cb}|,$ and $\gamma$.  They constrain $|V_{ub}/V_{cb}|$
by including the difference from its measured values, $0.08 \pm 0.02$, as a
term in $\chi^2$.

They fit 14 branching fractions: $K^- \pi^+,\ K^- \pi^0,\ \overline K^0
\pi^- ,$ $\overline K^0 \pi^0,$ $\pi^+ \pi^-,$ $\pi^- \pi^0,$ $\rho^0
\pi^-,$ $\omega \pi^-,$ $\rho^\mp \pi^\pm,\ K^{* -} \pi^+,\ \omega K^-,\
\omega \overline K^0,\ \phi K^-,$ and $\phi \overline K^0$. They
leave the $\eta^\prime$ and $\eta$ decay modes out of the fit, as something
strange is happening with these modes.  The $\chi^2$ of the fit, as a
function of $\gamma$, is shown in Fig. 12. The fit gives $\gamma = 114 \pm
23$ degrees.

\begin{figure}[h]
\begin{center} \epsfig{file=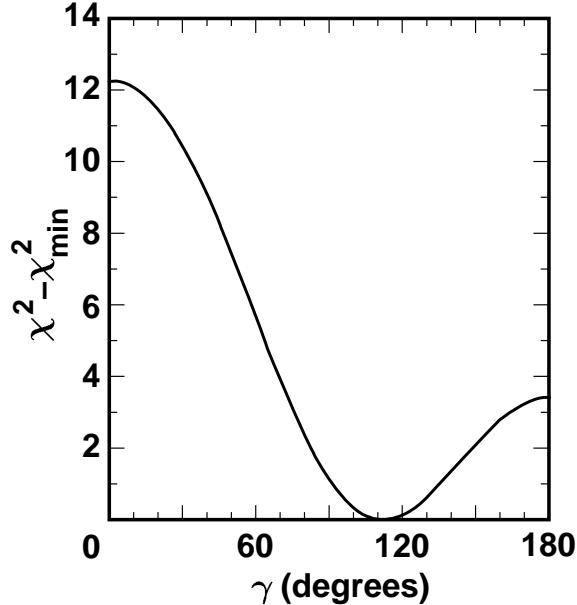, width=3in} \end{center}
\caption{$\chi^2$ of fit to 14 rare hadronic $B$ decay branching fractions,
as a function of $\gamma$, the phase of $V_{ub}^*$ (Hou-Smith-W\"urthwein).}
\label{fig:Hou-JGS-FKW}
\end{figure}

The error just quoted is that from the branching fraction errors only, and
does {\it not} include anything for theoretical uncertainty. Those must be
estimated and included before a serious number for $\gamma$ can be quoted.
However, from this exercise, so far, we can see that the data contain
information sufficient for a precise determination of $\gamma$, given
adequate theoretical understanding. Further, they argue for a large value
of $\gamma$. I'll take the liberty of assuming that the theoretical error
won't be more than $\sim \pm 50^\circ$, and interpret the rare $B$ results
as saying $\gamma > 60^\circ$.

\section{The Radiative Penguin Decay {\boldmath $b \rightarrow  s \gamma$}}

In Section 3.2.2, I argued that electroweak penguin processes, in
particular $b \rightarrow  s \gamma$, probe for New Physics up to masses
$\sim$500 GeV. What's been learned so far?

\subsection{The Exclusive Decay $B \rightarrow K^* (890) \gamma$}

The observation of $B \rightarrow  K^* (890) \gamma$, in 1993, was the
first clear observation of a penguin process. That analysis combined
conventional $B$ reconstruction techniques with continuum suppression
techniques, and used a likelihood ratio approach for further evidence.
While the existence of the radiative penguin process $b \rightarrow  s
\gamma$ was clearly established by this analysis, it did not provide a good
measurement of the inclusive rate (the theoretically interesting quantity),
since the theoretical estimates of $\Gamma (B \rightarrow  K^*
\gamma)/\Gamma (b \rightarrow  s \gamma)$ ranged from 5\% to 90\%. A direct
measurement of $b \rightarrow  s \gamma$ was called for.

\subsection{Branching Fraction for $b \rightarrow  s \gamma$}

The inclusive decay $b \rightarrow  s \gamma$ gives a monoenergetic photon
in the $b$ quark rest frame. That monoenergetic line is Doppler broadened
by the motion of the $b$ quark in the $B$ meson frame, and the motion of
the $B$ meson in the lab frame. But it remains a relatively narrow
distribution. In Fig. 13, I show the photon energy distribution expected from
$b \rightarrow s \gamma$, along with that expected from 
 other $B$ decay processes. The $b \rightarrow  s \gamma$ decays extend
beyond those from other $B$ decay processes and a study of the photon
spectrum above 2.0 GeV should cleanly give $b \rightarrow  s \gamma$.

\begin{figure}[h]
\begin{center} \epsfig{file=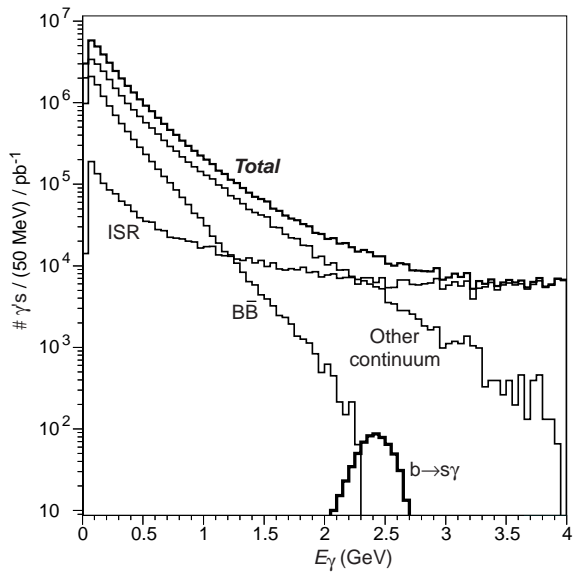, width=3in} \end{center}
\caption{The photon energy spectrum expected from $b \rightarrow s \gamma$;
other $B$ decay processes; initial state radiation $e^+ e^- \rightarrow
\gamma q \overline q$; and other continuum processes $e^+e^- \rightarrow q
\overline q \rightarrow \pi^0/\eta/\omega \rightarrow \gamma$. The sum of
ISR, other continuum processes and other $B$ decay processes is also shown.}
\label{fig:bsgam-spectra-expected}
\end{figure}

But wait. There are other curves shown on Fig. 13. One is the photon
spectrum from initial state radiation in continuum production, $e^+ e^-
\rightarrow  q \overline q \gamma$. The other, the spectrum of $\gamma$'s
from $\pi^0$ decay in continuum production, $e^+ e^- \rightarrow  q
\overline q \rightarrow  \pi^0 X \rightarrow  \gamma \gamma X$. The sum of
these two processes is more than two orders of magnitude larger than $b
\rightarrow  s \gamma$, at the $b \rightarrow  s \gamma$ peak. Continuum
suppression is absolutely essential.

In our 1995 measurement of the rate for $b \rightarrow  s \gamma$, we used
two different methods for continuum suppression. The first used eight
carefully chosen event-shape variables. While no individual variable has
strong discriminating power, each possesses some. We combined the eight
variables into a single variable $r$, which tends toward $+1$ for $b
\rightarrow  s \gamma$ and tends towards $-1$ for ISR and $q \overline q$.
We used a neural network for the task of combining the eight variables into
a single variable.  This was CLEO's first use of a neural network, and was
single handedly pushed through the collaboration by Jesse Ernst, against
strong opposition, much of it from his thesis advisor (me). That neural
networks are now used extensively, and intelligently, within CLEO can be
attributed to Jesse's good understanding of the strengths and limitations
of the technique.

The second method for continuum suppression has been dubbed \lq\lq
pseudoreconstruction''. In it, a high energy photon is combined with a kaon
($K^\pm$ or $K^0_s$) and 1-4 pions (of which one may be a $\pi^0$),
and tested for consistency with being a reconstructed $B$. (A $\chi^2$
composed of $B$ mass and $B$ energy, $\chi^2_B$, is used for this test.)
For those events with a pseudoreconstructed $B$, $\theta_{tt}$, the angle
between the thrust axis of the candidate $B$ and the thrust axis of the
rest of the event, gives additional discrimination against continuum
background. In pseudoreconstruction, often one does not have the totally
correct combination of particle (hence the ``pseudo''), but this is not
important (here), because the method is used only to suppress background,
and {\it not} for a mode-by-mode $B$ reconstruction analysis.

In our 1995 result, we performed two separate analyses, the event-shape
analysis and the pseudoreconstruction analysis, and averaged the branching
fractions obtained from each (allowing for a small amount of event
overlap). That result, ${\cal B} (b \rightarrow  s \gamma) = (2.32 \pm 0.57
\pm 0.35) \times 10^{-4}$, was based on a data sample of 3.0$fb^{-1}$.

More recently, we've combined the two continuum suppression techniques into
a single, unified analysis. For all events containing a high energy photon,
we compute the neural net variable $r$. For the subset of events that
pseudoreconstruct, with very loose requirements, we also calculate
$\chi^2_B$ and $\cos \theta_{tt}$. For these events, we feed $\chi^2_B$,
$\cos \theta_{tt}$, and $r$, into another neural network, obtaining a new
net variable $r_{\rm comb}$.  We assign a weight to each event, based on
$r_{\rm comb}$ for pseudoreconstructed events, and on $r$ for those events
which fail to reconstruct.  In this way we've analyzed a 4.7$fb^{-1}$ data
sample. The photon energy spectrum obtained is shown in Fig. 14. The
branching fraction obtained is 
$${\cal B} (b \rightarrow  s \gamma) = (3.15 \pm 0.35 \pm 0.32 \pm 0.26)
\times 10^{-4}$$
where the errors, in order, are statistical, systematic, and model
dependent. This number is in excellent agreement with the NLO prediction of
$(3.28 \pm 0.33) \times 10^{-4}$ of Chetyrkin, Misiak and M\"unz.

\begin{figure}[h]
\begin{center} \epsfig{file=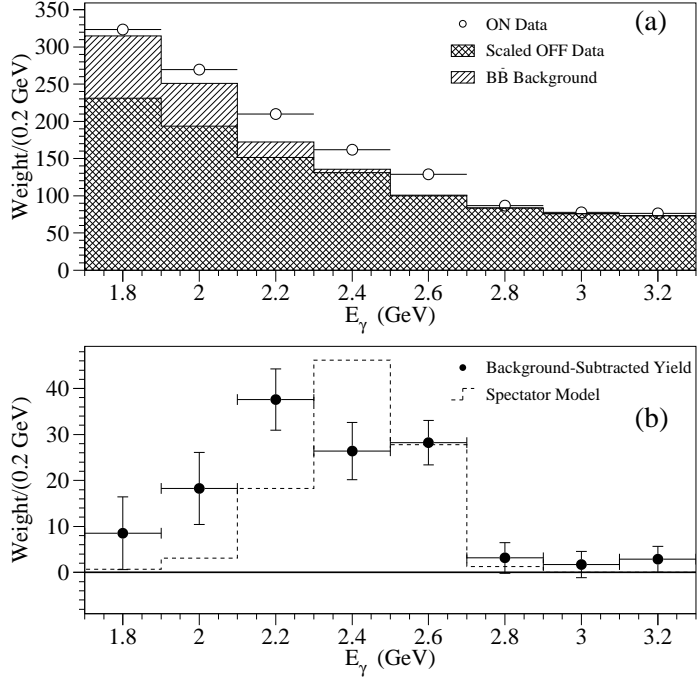, height=4in} \end{center}
\caption{The measured photon energy spectrum. (a) On-$\Upsilon$(4S) yield
(points); scaled off-$\Upsilon$(4S) yield (grey); contribution from $b$
decays other than $b \rightarrow s \gamma$ (black). (b) Net yield (points)
compared with spectator model prediction for $b \rightarrow s \gamma$.}
\label{fig:bsgam-measured}
\end{figure}

The comparison of experimental result with Standard Model prediction can be
(has been) used to place restrictions on New Physics. For example, our
conservative upper limit on the branching fraction, $4.5 \times 10^{-4}$,
rules out a charged Higgs with Model II coupling for Higgs masses less than
200 GeV. (In SUSY, there would be additional particles, which could
contribute with opposite sign, so the limitation is more complicated.
However, a hunk of SUSY parameter space is ruled out.)

CLEO now has 14$fb^{-1}$, 3 times the integrated luminosity used in the
analysis just described. What's holding us back? Well, look at the three
errors on the branching fraction. Reducing the statistical error by
$1/\sqrt{3}$ will do little good unless systematics and model dependence
can be beaten down. That takes more time.

\subsection{CP Asymmetry in $b \rightarrow  s \gamma$}

The CP asymmetry in $b \rightarrow  s \gamma$, ${\cal A}$, defined by
$${\cal A} \equiv {|A (b \rightarrow  s \gamma)|^2 - |A (\overline b
\rightarrow  \overline s \gamma )|^2 \over |A (b \rightarrow  s \gamma)|^2
+ |A (\overline b \rightarrow  \overline s \gamma )|^2} \quad ,$$
is very small, less than 1\%, in the Standard Model. So, observing a
nonzero value would be clear evidence for New Physics.

Suppose, in addition to the Standard Model decay amplitude for $b
\rightarrow  s \gamma$, $A_{SM}$, there is a New Physics amplitude, which
differs in weak phase from $A_{SM}$ by $\theta_W$, and in strong phase by
$\theta_S$. Then
$$A(b \rightarrow  s \gamma) = A_{SM} + A_{\rm New} e^{i \theta_S} e^{i
\theta_W}\quad ;$$
$$A(\overline b \rightarrow  \overline s \gamma) = A_{SM} + A_{\rm New}
e^{i \theta_S} e^{-i \theta_W}\quad .$$
The $(b/\overline b)$ averaged branching fraction, ${\cal B}$, is
$${\cal B} = {1 \over 2} [ |A(b  \rightarrow  s \gamma)|^2 
 + |A(\overline b  \rightarrow  
 \overline s \gamma)|^2] \approx A^2_{SM} (1 + 2 \rho \cos \theta_S \cos
\theta_W + \rho^2) \quad ,$$
where $\rho = A_{\rm New}/A_{SM}$.  The CP asymmetry ${\cal A} \approx 2
\rho \sin \theta_S \sin \theta_W$.

If one is sensitive to branching fraction differences of 20\%, then one can
detect New Physics amplitudes that are 10\% of the Standard Model
amplitude, if $\theta_W$ is near zero or 180 degrees, but cannot detect New
Physics amplitudes smaller than 45\% of the SM amplitude, if $\theta_W$ is
near $90^\circ$. For $\theta_W$ near $90^\circ$, ${\cal A} \approx 2 \rho
\sin \theta_S$.
So, if one were sensitive to CP asymmetries of 0.10, then one would have
sensitivity to this New Physics for $\rho \sin \theta_S > 0.05$.

So, there {\it is} a portion of New Physics parameter space, albeit small,
where New Physics will show up as a CP asymmetry, but not as a branching
fraction difference. This is discussed in general by A. Kagan and M.
Neubert (hep-ph/9803368), and as applied to SUSY by Aoki, Cho, and Oshimo
(hep-ph/9811251). Asymmetries in the 0.05-0.20 range are mentioned.

How might CLEO measure CP asymmetries in $b \rightarrow s \gamma$? By
pseudoreconstruction! But wait a minute, didn't I just say, in Section 6.2,
that ``In pseudoreconstruction, often one does not have the totally
correct combination of particles (hence the `pseudo'), but this is
not important, because the method is used only to suppress background
$\dots$''? Well, yes. It {\it still} isn't necessary to get the totally
correct combination of particles, but it {\it is} necessary to get the
flavor -- $b$ or $\overline b$ -- right. It turns out we get the flavor
right about 92\% of the time. It is straightforward to correct for the 8\%
mistake rate, a 19\% scaling up of the measured asymmetry. With the
4.7$fb^{-1}$ data sample used for the most recent branching fraction
analysis, we obtain a corrected asymmetry of
$${\cal A} = 0.16 \pm 0.14 \pm 0.05$$
So, no evidence for CP violation, but errors that are uncomfortably large.
The errors shown are statistical and systematic, in that order. With
relatively little work, the systematic error can be reduced substantially,
so even with 3 times the luminosity (our in-hand 14$fb^{-1}$), the
measurement will be statistics limited. An error of $\pm 0.08$ should be
straightforward to achieve. We're looking for ways to push that down,
giving consideration to lepton tagging as a possibility.

\section{Summary}

In ``the Early Days'', the basic features of the bottom quark were
established:

\begin{itemize}
\item A left-handed doublet with a very heavy top.

\vspace{-.1in}
\medskip
\item Decaying dominantly to charm, $b \rightarrow c W^-_V$. Coupling to
the second generation, $|V_{cb}| \approx 0.04$, smaller than the coupling
between second generation and first, $|V_{us}| = 0.22$.

\vspace{-.1in}
\medskip

\item Decay to up, $b \rightarrow u W^-_V$, suppressed relative to decay to
charm, but not zero.

\end{itemize}

In ``Recent Times'', the emphasis is on testing the Standard Model,
searching for New Physics. There are two approaches: measuring rates for
electroweak penguins, and ``overdetermining the CKM matrix''.  Lets see
where we now stand on each, and where we are going.

On electroweak penguins, the branching fraction for $b \rightarrow s
\gamma$ has been measured to $\pm$17\%, and is in good agreement with the
Standard Model. In the near future, with data already in hand, the accuracy
should be improved, to $\pm$10\%. At that point, the error on the
measurement will be about equal to the error on the theoretical prediction,
and further progress will be slower in coming.

The CP asymmetry has been measured to an accuracy of 
$\pm$0.14, and should soon improve to
$\pm$0.08. Further improvements are straightforward, as the error is purely
statistical, and the large data samples to be accumulated by BaBar, Belle,
and CLEO, in the next 3-4 years, should give a sensitivity to asymmetries
in the 0.05 range.

The electroweak process $b \rightarrow s \ell^+ \ell^-$ has not yet been
seen, though seeing it may not be far off. Very large data samples will be
required to study the various distributions that this 3-body final state
makes available. Possibly this is the role for hadron colliders.

\begin{figure}[th]
\begin{center} \epsfig{file=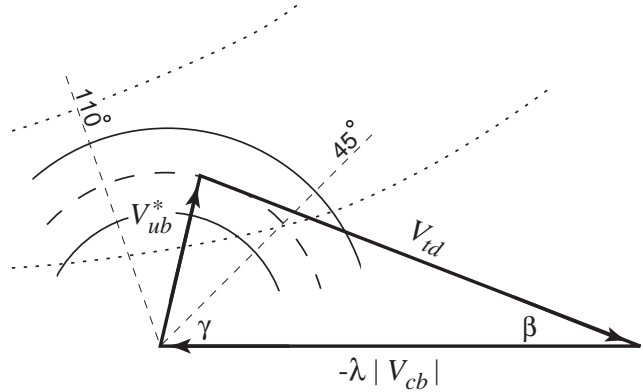, width=3.5in} \end{center}
\caption{Unitarity triangle, obtained from unitarity condition $V_{ub}^* -
\lambda |V_{cb}| + V_{td} =0$. Error band shown for $|V_{ub}|$ is
$\pm$25\%. The constraints $\gamma > 45^\circ$ (my interpretation of Hou,
Smith, W\"urthwein) and $\gamma < 110^\circ$ (limit on $B_s - \overline
B_s$ mixing) are indicated. The band allowed by $\epsilon$, the CP violating
parameter in neutral kaon decay, assuming a Standard Model, K-M origin, is also
shown.}
\label{fig:triangle}
\end{figure}

Concerning ``overdetermining the CKM matrix'', I would argue that the
CKM matrix has now been ``determined''. Figure 15 shows the famous
unitarity triangle, the one from the unitarity condition obtained by
multiplying the first column of the CKM matrix by the complex conjugate of
the third column:
$$V_{ud} V^*_{ub} + V_{cd} V^*_{cb} + V_{td}V^*_{tb} \approx V^*_{ub} -
\lambda |V_{cb}| + V_{td} = 0 \eqno(7\hbox{-}1)$$
The base of the triangle, $\lambda |V_{cb}|$, is known to $\pm$10\% (soon
to be $\pm$4\%). The left-hand leg, $|V_{ub}|$, is known to $\pm$25\%. The
one sigma error band for $|V_{ub}|$ is shown. The angle $\gamma$ probably
lies between $60^\circ$ and $90^\circ$, the lower limit coming from the
Hou, Smith, W\"urthwein analysis of CLEO data, the upper limit from $B_s -
\overline B_s$ mixing. I've shown more conservative limits on Fig. 15,
$45^\circ$ and $110^\circ$. That, I claim, ``determines'' the CKM
matrix. With that, one can predict the angle $\beta$ to be $20^\circ \pm
5^\circ \pm 2^\circ$, where the first error comes from the uncertainty in
$|V_{ub}|$, and the second from the uncertainty in $\gamma$. (Note that the
uncertainty in the prediction of $\beta$ comes dominantly from $|V_{ub}|$,
not $\gamma$.)

The first ``overdetermination'' of the CKM matrix comes from the CP
violating parameter $\epsilon$ in neutral kaon decay. The band it defines
nicely intersects the allowed region.

The next ``overdetermination'' will be BaBar and Belle's measurements
of $\beta$. With 30$fb^{-1}$ data samples, they expect to measure $\beta$
to $\pm 5^\circ$. It will be interesting to see how their results compare
with CLEO's prediction of ($20 \pm 5)^\circ$, and also how their error on
measured $\beta$ will compare with CLEO's predictions based on improved
measurement of $|V_{ub}|$. Interesting times ahead!

\section{Acknowledgements}

I have benefitted immeasurably from countless interactions and discussions with
my collaborators in CLEO over the past two decades.  I wish to thank them for
this, and absolve them of any blame for my rash statements in this
paper.

\end{document}